\documentclass{jfm}
\usepackage{amssymb}
\usepackage[dvips]{graphicx}

\ifCUPmtlplainloaded \else
  \checkfont{eurm10}
  \iffontfound
    \IfFileExists{upmath.sty}
      {\typeout{^^JFound AMS Euler Roman fonts on the system,
                   using the 'upmath' package.^^J}%
       \usepackage{upmath}}
      {\typeout{^^JFound AMS Euler Roman fonts on the system, but you
                   dont seem to have the}%
       \typeout{'upmath' package installed. JFM.cls can take advantage
                 of these fonts,^^Jif you use 'upmath' package.^^J}%
      }
  \else
  \fi
\fi

\ifCUPmtlplainloaded \else
  \checkfont{msam10}
  \iffontfound
    \IfFileExists{amssymb.sty}
      {\typeout{^^JFound AMS Symbol fonts on the system, using the
                'amssymb' package.^^J}%
       \usepackage{amssymb}%
       \let\le=\leqslant  
       \let\ge=\geqslant  
      }{}
  \fi
\fi

\ifCUPmtlplainloaded \else
  \IfFileExists{amsbsy.sty}
    {\typeout{^^JFound the 'amsbsy' package on the system, using it.^^J}%
     \usepackage{amsbsy}}
    {\providecommand\boldsymbol[1]{\mbox{\boldmath $##1$}}}
\fi

\newsavebox{\astrutbox}
\sbox{\astrutbox}{\rule[-5pt]{0pt}{20pt}}

\newcommand\etal{\mbox{\textit{et al.}}}

\title[Linear stability analysis of 
granular flow on a slope]
{Linear stability analysis of rapid granular flow 
on a slope and density wave formation}

\author[N. Mitarai and H. Nakanishi]%
{N\ls A\ls M\ls I\ls K\ls O\ns M\ls I\ls T\ls A\ls R\ls A\ls I\ns 
\and H\ls I\ls I\ls Z\ls U\ns N\ls A\ls K\ls A\ls N\ls I\ls 
S\ls H\ls I}

\affiliation{Department of Physics, Kyushu University 33,
Fukuoka 812-8581, Japan
}

\pubyear{??}
\volume{??}
\pagerange{??--??}
\date{\today} 
\begin{document}

\maketitle

\begin{abstract}
The linear stability of rapid granular flow on a slope under gravity
against the longitudinal perturbation
is analyzed using hydrodynamic equations.
It is demonstrated that 
the steady flow uniform along the flow direction becomes
unstable against the long-wavelength perturbations
longitudinal to the flow direction
for certain parameter ranges to form the density wave,
in contrast with the finite wavelength instability
against the transverse perturbation 
(\cite[Forterre \& Pouliquen 2002]{FP02}).
It is shown that the instability can be understood 
as the the long-wave instability
of the kinematic waves in a quasi-one dimensional system.
The results are compared 
with our previous molecular dynamics simulations
(\cite[Mitarai \& Nakanishi 2001]{MN01}),
where the spontaneous density wave formation has been found.
\end{abstract}

\section{Introduction}\label{intro}
Granular flow exhibits a variety of dynamical phenomena,
which have been attracting research interests for years 
(for reviews, see e.g.
\cite[Savage 1984]{S84} and 
\cite[Jaeger, Nagel \& Behringer 1996]{JNB96}). 
Its complex behaviors can be seen
even in a simple situation like the 
gravitational flow on a slope.
When the inclination angle is large and the slope
is rough, a rapid and relatively low-density flow
is realized, and 
the interaction between grains is dominated by 
inelastic collisions. 
On the other hand, when the inclination angle is small, 
the flow becomes dense and slow, and the frictional 
interaction plays an important role 
(\cite[Savage 1984]{S84}; 
\cite[Mitarai \& Nakanishi 2003]{MN03}).
The comprehensive rheology of the granular flow
has not been established yet,
except for the rapid collisional flow regime, 
where the hydrodynamic models 
have been developed with the
constitutive relations based on
the kinetic theory of inelastic hard spheres
(\cite[Jenkins \& Savage 1983]{JS83};
\cite[Campbell 1990]{C90};
\cite[Lun \etal 1984]{LSJC84};
\cite[Goldhirsh 2003]{G03});
it has been demonstrated that a certain quantitative
agreement can be achieved for the steady flow
by introducing the spinning motion of each grains
(\cite[Mitarai, Hayakawa \& Nakanishi 2002]{MHN02}).
The steady granular flow, however, turns out to be unstable in various
ways, and shows rich phenomena.

In the experiment on a shallow granular flow on a wide slope,
\cite{FP01} have observed that there appears a regular
pattern of longitudinal streaks along the flowing direction.
This phenomenon has been analyzed by means of the hydrodynamic equations
for rapid granular flow (\cite[Forterre \& Pouliquen 2002]{FP02}).
They have calculated the steady solutions
and examined its linear stability numerically.
They have found that, at a certain parameter region,
the steady flow shows the ``inverted density profiles'',
in which the maximum density appears not at the bottom
but at a finite distance from the bottom because of the agitation 
by the collisions with the rough solid bottom.
It has been shown that the solutions with 
the ``inverted density profiles'' are 
unstable against the perturbations
transverse to the flowing direction,
and  the instability results in 
the vortex patterns analogous to 
the rolls in the Rayleigh-B\'ernard convection;
the streaks found in the experiment were interpreted as the 
result of the rolls of vortices.

Another instability that has been observed is
the density wave formation along the flowing direction;
experimentally, \cite{LK01} have observed
the jamming patters traveling
upstream in the dense chute flow,
and \cite[Prasad, Pal \& R\"omkens (2000)]{PPR00}
have found the waves develops in the shallow flow 
as they travel downstream. 
The present authors have performed the molecular dynamics simulations 
of two-dimensional granular flow on a slope 
and found that the steady flows are unstable against the density wave
formation when the length of the slope is long enough 
and/or the particle density is low enough
(\cite[Mitarai \& Nakanishi 2001]{MN01}) .

The purpose of this paper is to perform the 
linear stability analysis on the hydrodynamic equations
to investigate the nature of the density wave formation 
instability found in the experiments and the numerical simulations.
The basic method is the same with that
used in \cite{FP02},
but we examine the stability against the perturbations
longitudinal to the flowing direction 
in contrast with the case of \cite{FP02}, where the transverse
stability has been studied.

This paper is organized as follows.
In \S \ref{hydromodel}, 
the hydrodynamic model for rapid granular flow 
is introduced.
The steady solutions uniform along the flow direction
are numerically obtained
in \S \ref{secsteady},
and the results of the linear stability analysis is presented
in \S \ref{lstaba}.
In \S \ref{discussion}, discussions 
and the comparison with the molecular dynamics simulations are given.
The results are summarized in \S \ref{sum}.

\section{Hydrodynamic Equations for Granular Flows}\label{hydromodel}
\subsection{Hydrodynamic equations and constitutive relations}
The hydrodynamic fields for granular flows in three dimensions
are the mass density $\rho$,
the mean velocity $\boldsymbol u$, and the granular temperature
$T$, where $T=<\delta \boldsymbol u^2>/3$. 
Here, $\delta \boldsymbol u=\boldsymbol u -<\boldsymbol u>$ 
and $<\dots>$ represents the average over the microscopic scale. 
Under gravity, they follow
\begin{eqnarray}
\left(\frac{\partial}{\partial t}+\boldsymbol{u}\cdot\nabla\right)
\rho&=&-\rho\nabla \cdot \boldsymbol u,
\label{eq:rho}\\
\rho\left(\frac{\partial}{\partial t}+\boldsymbol{u}\cdot\nabla\right)
\boldsymbol u&=&\rho \boldsymbol g
-\nabla \cdot\Sigma,
\label{eq:u}\\
\frac{3}{2}\rho\left(\frac{\partial}{\partial t}+\boldsymbol{u}\cdot\nabla\right)
T&=&-\nabla \cdot \boldsymbol q
-\Sigma : \nabla \boldsymbol u-\Gamma,
\label{eq:temp}
\end{eqnarray}
with the acceleration of gravity $\boldsymbol g$, 
the stress tensor $\Sigma$, the heat flux $\boldsymbol q$,
and the energy loss $\Gamma$ due to the 
inelastic nature of interactions between grains.

We employ the constitutive relations
derived by \cite[Lun \etal (1984)]{LSJC84} for three dimensional system
based on the kinetic theory of the inelastic particles:
\footnote{
The original form of $\boldsymbol q$
derived by \cite[Lun \etal (1984)]{LSJC84} is
$\boldsymbol q=-\kappa \nabla T-\kappa_h \nabla \nu$.
The coefficient $\kappa_h$ is proportional to $(1-e_p)$, 
thus disappears in the elastic limit.
We checked that the influence of the term 
$\kappa_h \nabla \nu$ on the steady solutions is small
in the considered parameter region, 
therefore neglected this term as \cite{FP02}.}
\begin{eqnarray}
\Sigma&=&(p-\zeta \nabla \cdot \boldsymbol u)I
-2\mu S,\\
\boldsymbol q&=&-\kappa \nabla T,
\end{eqnarray}
where
\begin{equation}
S=\frac{1}{2}[\nabla \boldsymbol u+(\nabla \boldsymbol u)^t]-\frac{1}{3}
(\nabla\cdot\boldsymbol u)I,
\end{equation}
and 
\begin{eqnarray}
&& p(\nu,T)=\rho_p f_1(\nu)T, \quad \mu(\nu,T)=\rho_p \sigma f_2(\nu)T^{1/2},
\quad \zeta(\nu,T)=\rho_p \sigma f_3(\nu)T^{1/2},\nonumber \\
&&\kappa(\nu,T)=\rho_p \sigma f_4(\nu)T^{1/2}, \quad 
\Gamma(\nu,T)=\frac{\rho_p}{\sigma}(1-e_p^2)f_5(\nu)T^{3/2},
\label{eq:pmuetc}
\end{eqnarray}
with the material density of particle $\rho_p$,
the packing fraction $\nu=\rho/\rho_p$,
the particle diameter $\sigma$,
and the restitution coefficient between particles $e_p$.
Here, $I$ represents the unit matrix.
The dimensionless functions $f_i(\nu)$ ($i=1,\dots,5$)
are given in Table \ref{table}.
For the radial distribution function $g_0(\nu)$ in
these functions, we adopted the form suggested by \cite{LS86}:
\begin{equation}
g_0(\nu)=\frac{1}{(1-\nu/\nu_m)^{2.5\nu_m}},
\end{equation}
with the maximum solid fraction $\nu_m$, for
which we use $0.60$ as \cite{FP02}. 

In the following, all the variables
are non-dimensionalized 
by the length unit $\sigma$, 
the mass unit $\rho_p \sigma^3$,
and the time unit $\sqrt{\sigma/g}$.
The density field is expressed by
the packing fraction $\nu$ instead of the mass density $\rho$. 
The restitution coefficient 
between particles is set to be $e_p=0.7$, that is the value 
used in our previous simulations
(\cite[Mitarai \& Nakanishi 2001]{MN01}).

\begin{table}
\begin{tabular}{l}
$\displaystyle
f_1(\nu)=\nu(1+4\eta \nu g_0(\nu))
$\\ \\
$\displaystyle
f_2(\nu)=\frac{5\pi^{1/2}}{96\eta(2-\eta)}
\left(1+\frac{8}{5}\eta\nu g_0(\nu)\right)
\left(\frac{1}{g_0(\nu)}+\frac{8}{5}\eta(3\eta-2)\nu\right)
+\frac{8}{5\pi^{1/2}}\eta\nu^2 g_0(\nu)
$\\ \\
$\displaystyle
f_3(\nu)=\frac{8}{3\pi^{1/2}}\eta\nu^2 g_0(\nu)
$\\ \\
$\displaystyle
f_4(\nu)=\frac{25\pi^{1/2}}{16\eta(41-33\eta)}
\left(1+\frac{12}{5}\eta\nu g_0(\nu)\right)
\left(\frac{1}{g_0(\nu)}+\frac{12}{5}\eta^2(4\eta-3)\nu\right)
+\frac{4}{\pi^{1/2}}\eta\nu^2 g_0(\nu)
$\\ \\
$\displaystyle
f_5(\nu)=\frac{12}{\pi^{1/2}}\nu^2 g_0(\nu)
$\\ \\
$\displaystyle
f_6(\nu)=\frac{\sqrt{3}\pi\nu g_0(\nu)}{2\nu_mf_4(\nu)}
$\\ \\
$\displaystyle
f_7(\nu)=\frac{\pi\nu g_0(\nu)}{2\sqrt{3}\nu_mf_2(\nu)} 
$\\ \\
\end{tabular}
\caption{Dimensionless functions used in the 
constitutive relations and the boundary conditions
with $\eta\equiv (1+e_p)/2$.}
\label{table}
\end{table}

\subsection{Boundary conditions}
The granular flow has a non-zero slip velocity at the solid boundary, 
where we should impose the momentum and the kinetic energy balances.
Sophisticated boundary conditions have been proposed based on
microscopic calculations of the 
kinetic theory for specific geometries
(\cite[Jenkins \& Richman 1986]{JR86};
\cite[Richman 1988]{R88};
\cite[Jenkins 1992]{J92}).
We employ, however, a simpler form of the boundary condition obtained
from 
a heuristic approach
(\cite[Johnson \& Jackson 1987]{JJ87};
\cite[Johnson, Nott \& Jackson 1987]{JNJ90});
\begin{eqnarray}
-\boldsymbol n \cdot \Sigma\cdot \boldsymbol t&=&\eta^*(\nu, T)|
\boldsymbol u_s|,
\label{eq:bc1}\\
\boldsymbol n\cdot \boldsymbol q&=&-\boldsymbol u_s\cdot\Sigma\cdot 
\boldsymbol n-\Gamma^*(\nu,T),
\label{eq:bc2}
\end{eqnarray}
where the unit vector $\boldsymbol n$ is normal to
the floor, $\boldsymbol u_s$ is the slip velocity at the floor,
and $\boldsymbol t=\boldsymbol u_s/|\boldsymbol u_s|$ is the unit vector 
along the slip velocity.

The first equation (\ref{eq:bc1}) expresses that
the stress at the boundary balances with
the momentum transfer due to the collisions
between the slope and the flowing grains.
The momentum transfer, or RHS of (\ref{eq:bc1}), 
is assumed to be given by
\begin{equation}
\eta^*(\nu,T)|\boldsymbol u_s|
=\frac{\pi}{6}\phi |\boldsymbol u_s|\Omega(\nu,T),
\end{equation}
with the collision rate $\Omega(\nu,T)$
per unit time per unit area.
Here, the factor $\pi/6$ comes from the 
non-dimensionalization of the particle mass $m=\sigma^3 \pi/6$.
The parameter $\phi $ characterizes the roughness of the
boundary, and the expression means that the
fraction $\phi $ of the particle momentum
is transfered to the boundary in each collision, therefore,
the larger value of $\phi $ represents the rougher boundary.
For the rough boundary in the two-dimensional system
with closed packed disks, 
\cite{JR86} estimated
as $\phi \approx 0.1$, but
\cite{FP02} adopted the smaller value $\phi =0.05$ 
for most of the cases because they expected that
a boundary with closed packed spheres in three-dimensional system
is smoother on average.
In this paper, we mainly use $\phi =0.05$,
but the case of $\phi =0.10$ is also examined
in order to see general tendencies.

The second equation (\ref{eq:bc2}) represents the energy balance, 
and means that the heat flux 
at the boundary comes from two effects, namely, 
the frictional heating due to the non-zero
slip velocity and the energy loss due to the
inelastic collision with the floor.
The energy loss term $\Gamma^*$ in (\ref{eq:bc2}) is given by
\begin{equation}
\Gamma^*= \Phi\cdot \frac{\pi}{6}\cdot \frac{3}{2} T \cdot \Omega(\nu,T) 
\label{eq:gammas}
\end{equation}
with the collision rate $\Omega(\nu,T)$.
The parameter $\Phi$ represents 
the rate of energy loss per collision\footnote{
\cite{JJ87} explicitly relate $\Phi$
to the restitution coefficient between the floor 
and the particles $e_w$ in the form $\Phi=(1-e_w^2)$,
but we adopt (\ref{eq:gammas}) as
a more general expression.
},
and we use $\Phi=0.39$ in this paper.

For the collision rate $\Omega(\nu,T)$, 
we use the form adopted by \cite{FP02}, i.e.,
$\Omega(\nu,T)=\sqrt{3T}\nu g_0(\nu)/\nu_m$:
then the expression of $\eta^*(\nu,T)$ and $\Gamma^*(\nu,T)$ 
in non-dimensionalized form are 
\begin{eqnarray}
\eta^*(\nu,T)&=&\phi f_7(\nu)f_2(\nu)T^{1/2},
\label{etastar} \\
\Gamma^*(\nu,T)&=&\frac{1}{2}
\Phi f_6(\nu)f_4(\nu)T^{3/2},
\label{gammastar} 
\end{eqnarray}
where the dimensionless functions 
$f_6(\nu)$ and $f_7(\nu)$ are given in Table \ref{table}.

\begin{figure}
\centerline{
\includegraphics[width=0.3\textwidth]{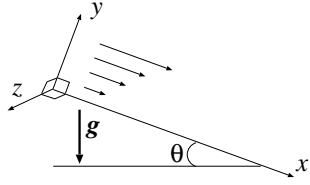}}
\caption{The coordinate system.}
\label{slope}
\end{figure}

At infinity,
we impose the condition that 
the stress and the heat flux vanish, 
namely,
\begin{equation}
\Sigma \to 0 \quad \mbox{and} \quad \boldsymbol{q}\to 0
\quad \mbox{as}\quad y\to \infty, 
\label{eq:inftycond}
\end{equation}
where 
the $y$ axis is taken perpendicular to the floor (figure \ref{slope}).

\section{Steady Flows}\label{secsteady}
\subsection{Equations and Numerical method}
First, we consider the steady solution which is uniform along the slope
for (\ref{eq:rho}) - (\ref{eq:temp})
with the boundary conditions (\ref{eq:bc1}), (\ref{eq:bc2}),
and (\ref{eq:inftycond}) in the form
\begin{eqnarray}
\nu(x,y,z,t)&=&\nu_0(y),\label{eq:rho0}\\
\boldsymbol u(x,y,z,t)&=&(u_0(y),0,0),\label{eq:u0}\\
T(x,y,z,t)&=&T_0(y),\label{eq:T0}
\end{eqnarray}
where we take the $x$ axis along the slope,
the $y$ axis perpendicular to the floor,
and the $z$ axis perpendicular to the $x-y$ plane 
(figure \ref{slope}).

Then,
the equations (\ref{eq:rho}) - (\ref{eq:temp}) are written as 
\begin{eqnarray}
0
&=&\nu_0 \sin\theta-\frac{\mbox{d}\Sigma_{xy}^0}{\mbox{d}y},
\label{eq:steady1}\\
0&=&-\nu_0 \cos\theta-\frac{\mbox{d}\Sigma_{yy}^0}{\mbox{d}y},
\label{eq:steady2}\\
0&=&
 -\Sigma_{xy}^0\frac{\mbox{d}u_0}{\mbox{d}y}
-\frac{\mbox{d}q_{y}^0}{\mbox{d}y}-\Gamma^0,
\label{eq:steady3}
\end{eqnarray}
where the superscript $0$ 
denotes that the functions are 
for the steady solution,
namely, 
$\boldsymbol{q}^0=-\kappa(\nu_0,T_0)\nabla T_0=(0,q_{y}^0,0)$, 
etc.
By integrating (\ref{eq:steady1})
and (\ref{eq:steady2}) over $y$
with the stress free condition at infinity,
we obtain the condition 
\begin{equation}
\Sigma_{xy}^0=-\tan\theta \ \Sigma_{yy}^0.
\label{eq:steady4}
\end{equation}
From (\ref{eq:steady1})-(\ref{eq:steady4})
and the constitutive relations, we have
\begin{eqnarray}
\nu_0'(y)&=&-\frac{f_1^0T_0'+\nu_0\cos\theta}{f_{1,\nu}^{0}T_0},
\label{eq:nu0}\\
u_0'(y)
&=&\frac{f_1^0T_0^{1/2}}{f_2^0}\tan\theta,
\label{eq:u00}\\
T_0''(y)
&=&\frac{1}{f_4^0}
\left[
(1-e_p^2)f_5^0T_0-f_2^0u'_{0}-f_{4,\nu}^{0} \nu'_{0}T'_{0}
-\frac{(T'_{0})^2}{2T_0}f_4^0
\right],
\label{eq:T00}
\end{eqnarray}
where $f_i^0\equiv f_i(\nu_0)$,
$f_{i,\nu}^0\equiv \frac{\mbox{d}}{\mbox{d}\nu}
f_i(\nu)|_{\nu=\nu_0}$,
and the prime indicates
the derivative by its argument.

The boundary conditions 
(\ref{eq:bc1}) and (\ref{eq:bc2})
at the floor ($y=0$)
for the steady solution can be written as
\begin{eqnarray}
T_0&=&\left(\frac{\phi  f_2^0 f_7^0}{f_1^0\tan\theta}
u_0\right)^2,
\label{eq:T0b}\\
T_0'&=&-f_6^0\left(
\frac{1}{3}\phi u_0^2
-\frac{1}{2}
\Phi T_0
\right).
\label{eq:dT0b}
\end{eqnarray}
The boundary condition (\ref{eq:inftycond})
that the stress and the energy flux should vanish at infinity 
is satisfied when (\cite[Ahn, Brennen \& Sabersky 1992]{ABS92})
\begin{equation}
T_0'(y)\to 0 \quad \mbox{when}
\quad y\to \infty.
\label{eq:bcinf}
\end{equation}

In order to obtain the steady solutions, we integrate
(\ref{eq:nu0}), (\ref{eq:u00}), and (\ref{eq:T00}) numerically 
using the fourth-order Runge-Kutta method
with the boundary conditions (\ref{eq:T0b}) and (\ref{eq:dT0b}).
We employ the shooting method to find
the solution which satisfies the condition (\ref{eq:bcinf}) 
(\cite[Forterre \& Pouliquen 2002]{FP02}); 
for a given inclination angle $\theta$ 
and a given density at the floor $\nu_0(0)$,
we search for a solution by adjusting the value of 
the velocity at the floor $u_0(0)$.
In the actual calculations, we integrate the equations
numerically from $y=0$ to a certain height
$y_{\max}$, and search for the solution
which gives $|T_0'(y_{\max})|< 10^{-7}$.
The value of $y_{\max}$ is chosen to be large enough
in comparison with the relaxation length, that
depends on the parameters and can be determined only after
the solution is obtained.

We use $\theta$ and $\nu_0(0)$ 
to specify the solution in the rest of the paper.

\subsection{Numerical solutions}\label{steadysol}
For a given roughness $\phi$ of the slope,
steady solutions are found for a certain range of
the inclination angle of the slope $\theta$ 
(\cite[Forterre \& Pouliquen 2002]{FP02}).
We present the steady solutions for the two cases, 
(i) $\phi$=0.05 and (ii) $\phi=0.10$;
the most of the results are for the case of $\phi=0.05$,
and the case of $\phi=0.10$ will be given to examine
general trend.

\subsubsection{The case of $\phi=0.05$}
\begin{figure}
\begin{minipage}{0.5\textwidth}
\includegraphics[width=0.8\textwidth]{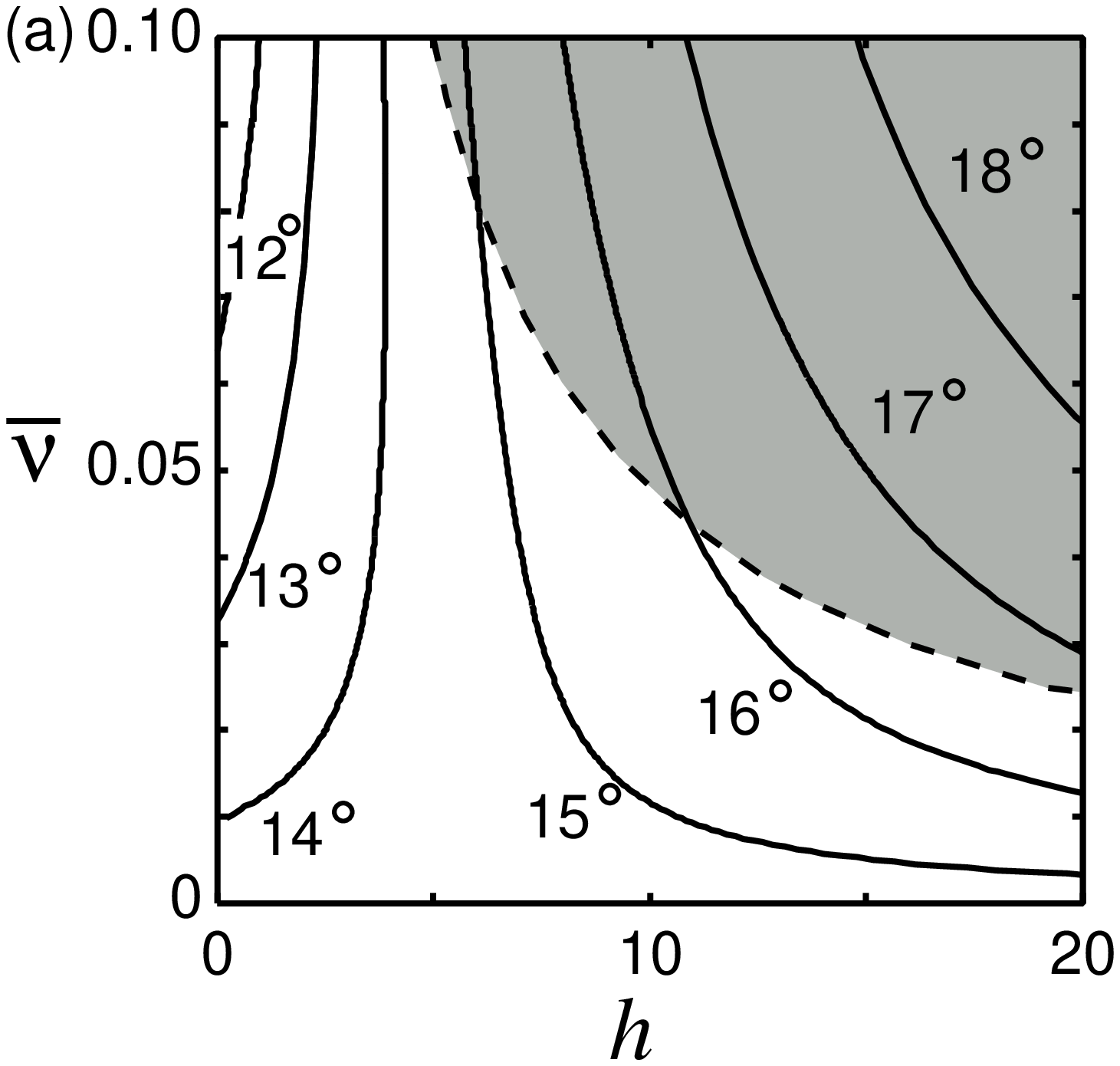}
\end{minipage}
\begin{minipage}{0.5\textwidth}
\includegraphics[angle=-90,width=\textwidth]{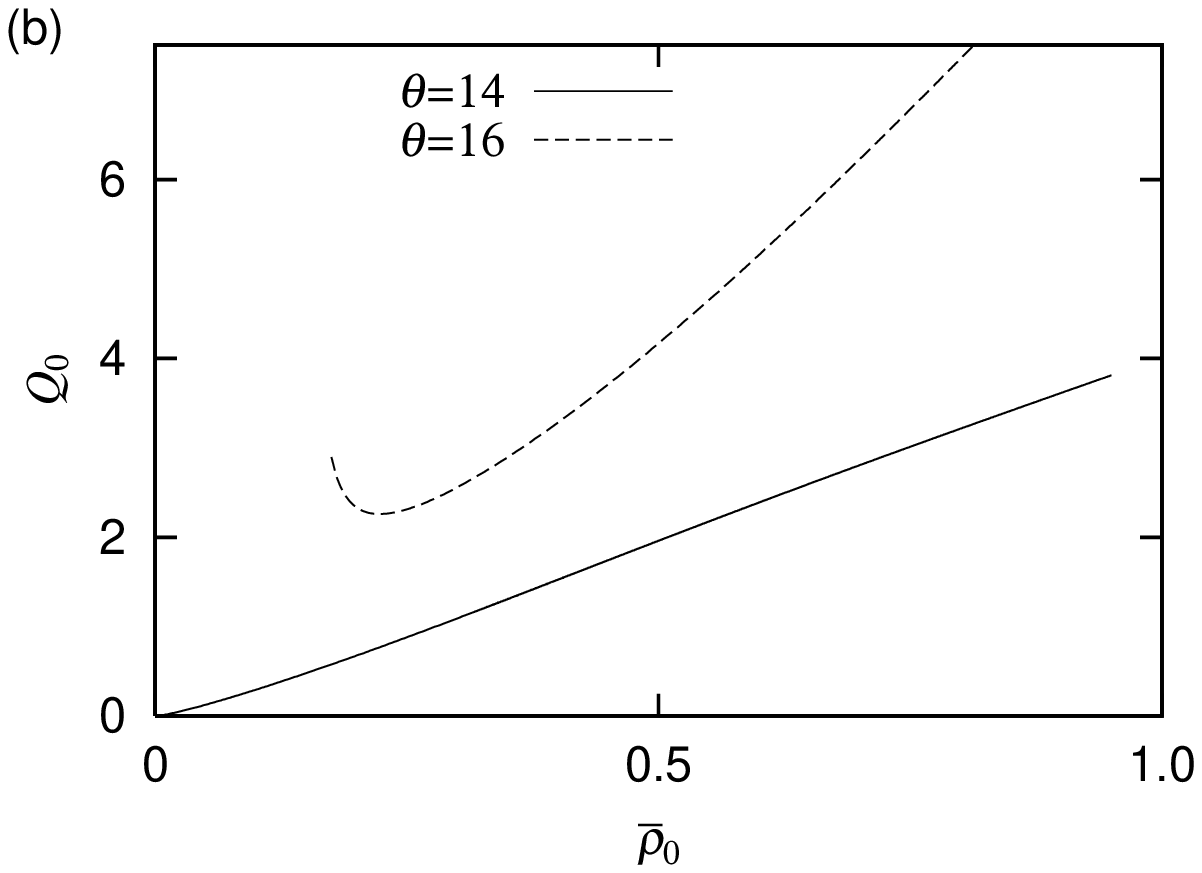}
\end{minipage}
\caption{(a)The contour lines of $\theta$
in the $(h,\bar \nu)$ plane for $\phi=0.05$.
The region of the non-monotonic density profiles
is shown by a grey region.
(b)The flux $Q_0$ vs 
the one-dimensional density $\bar \rho_0$ for $\phi=0.05$.
The solid line and the dashed line
are for $\theta=14^\circ$ and $\theta=16^\circ$,
respectively.
}
\label{steadyphase005}
\end{figure}
\begin{figure}
\includegraphics[angle=-90,width=0.32\textwidth]{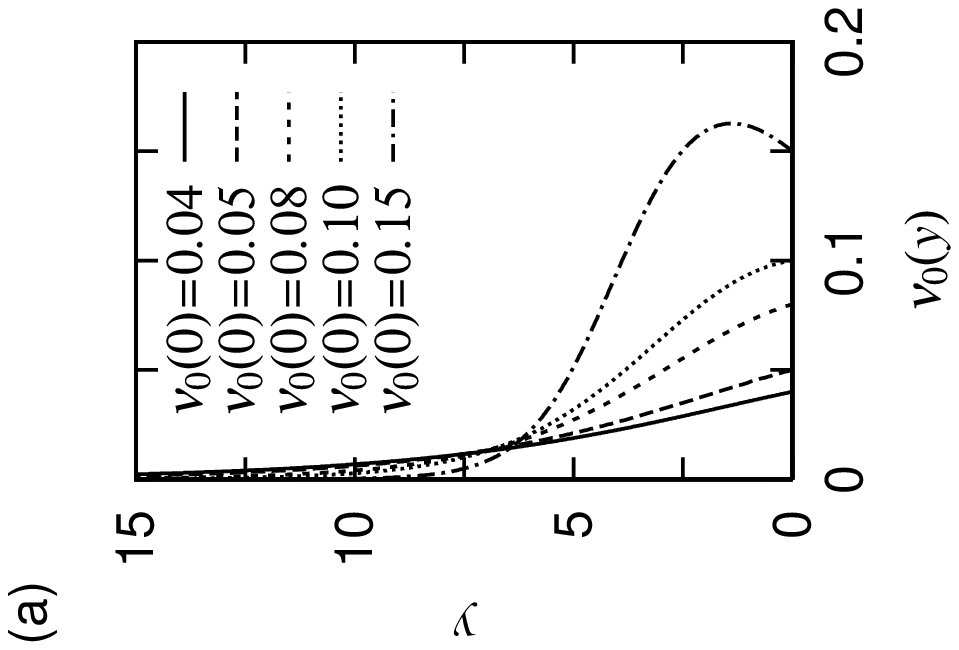}
\includegraphics[angle=-90,width=0.32\textwidth]{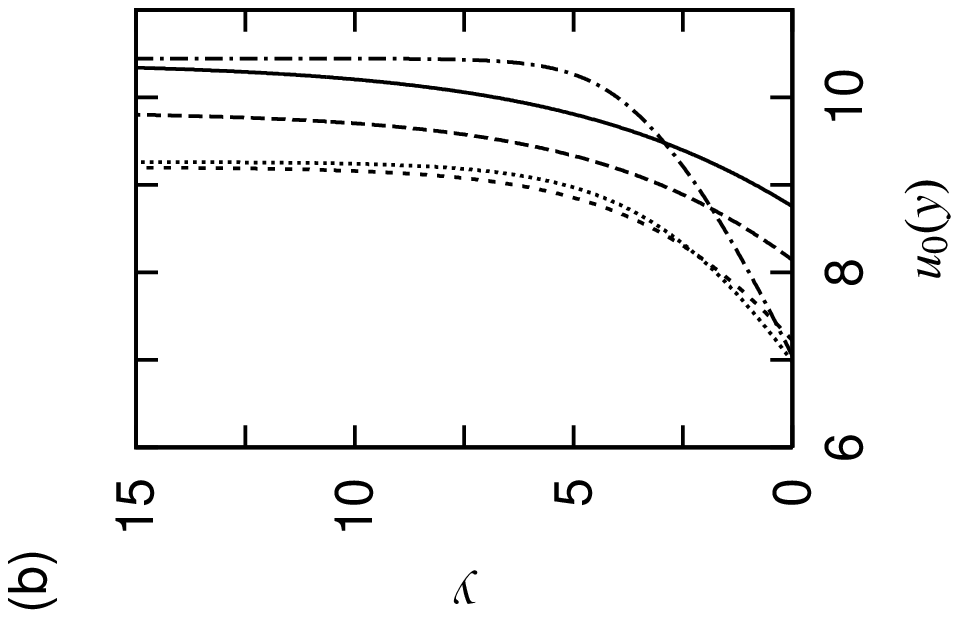}
\includegraphics[angle=-90,width=0.32\textwidth]{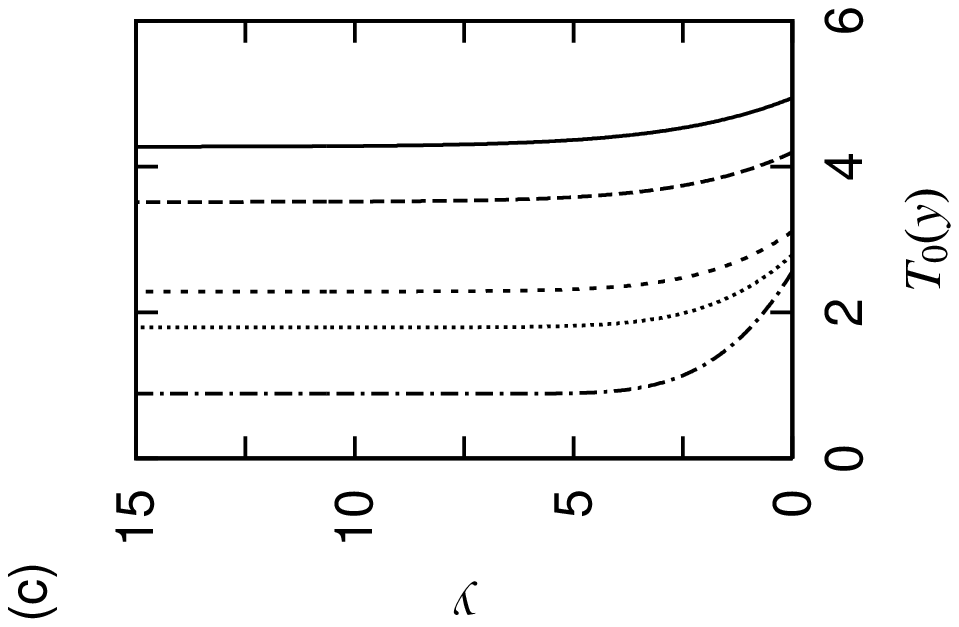}
\caption{The steady solutions for $\phi=0.05$ and $\theta=16^{\circ}$;
the density profiles (a), the velocity profiles (b),
and the temperature profiles (c).
Different line types correspond to the solutions with different
densities; $\nu_0(0)=0.04$ (solid lines),
$0.05$ (long-dashed lines),
$0.08$ (short-dashed lines), 
$0.10$ (dotted lines), 
and $0.15$ (dash-dotted lines).
}
\label{steady}
\end{figure}
For an appropriate  $\theta$, there exist steady solutions 
for a given density at the floor $\nu_0(0)$.
We numerically find the steady solution for the range
$8^{\circ}\lesssim \theta \lesssim 20^{\circ}$
for the moderate density; 
towards the lower limit of $\theta$, the length scale
of the density decay 
in the $y$-direction tends to zero
and becomes smaller than the particle diameter, 
which is physically unacceptable, 
while the decay length of density diverges towards
the upper limit of $\theta$.
This is consistent with the analysis by 
\cite{AJ92} that the steady solution 
in the high-density limit
is allowed for a finite range of $\theta$.

For a given steady solution, 
we define the  ``thickness'' $h$ 
and the ``mean density'' $\bar \nu$:
the thickness $h$ is the value of
$y$ where the density is 1\% of the maximum density,
and the mean density $\bar \nu $ is given by
\begin{equation}
\bar \nu=\frac{1}{h}\int^\infty_0\nu_0(y)\mbox{d}y.
\end{equation}
It has been found numerically by \cite{FP02}
that there exists a one to one correspondence 
between $(\theta, \nu_0(0))$ and $(h,\bar \nu)$.

Figure \ref{steadyphase005}(a) shows the contour lines of $\theta$
in the $(h,\bar \nu)$ plane, where $\theta$ increases
from left to right. In this plot, it is clear that
$h$ goes to zero as $\theta$ decreases 
and $h$ diverges as $\theta$ increases.
There is a separatrix near the bottom at $h\approx 7$,
where the value of $\theta$ is around $15^{\circ}$;
the contour line for $\theta\le 14^{\circ}$
leads to $h=0$ as $\bar \nu$ decreases,
while $h$ diverges along the contour lines
for $\theta\ge 15^{\circ}$ as
 $\bar \nu$ becomes small.

The difference between $\theta \ge 15^\circ$ 
and $\theta \le 14^\circ$ can be seen 
in the relation between the flux $Q_0$ and the one-dimensional density
$\bar \rho_0$ defined as
\begin{equation}
Q_0 \equiv \int^{\infty}_{0}\nu_0(y) u_0(y)\mbox{d}y
\quad \mbox{and} \quad
\bar \rho_0 \equiv\int^{\infty}_{0}\nu_0(y)\mbox{d}y,
\label{rho0q0}
\end{equation}
respectively, for a fixed inclination angle. 
It is found that $Q_0$ is an increasing function of $\bar \rho_0$ for
$\theta\le 14^\circ$, while $Q_0$ has a minimum at a finite 
$\bar \rho_0$ for $\theta\ge 15^\circ$;
the plot of $Q_0$ vs $\rho_0$ is shown in figure \ref{steadyphase005}(b)
for $\theta=14^\circ$ and $\theta=16^\circ$.

The typical profiles of the density, the velocity, 
and the temperature for $\theta=16^\circ$ are shown in figure \ref{steady}
for the density at the floor $\nu_0(0)=0.04\sim 0.15$.
We see in figure \ref{steady}(a) that
the density decays monotonically
when the density at the floor $\nu_0(0)$
is small enough ($\nu_0(0)\le 0.10$), while 
for higher density ($\nu_0(0)=0.15$)
the maximum density appears at a finite height.
The region where
the maximum density appears at a finite height
is shown in figure \ref{steadyphase005}(a) 
as a grey region.
We focus on the lower density region because
the density decays monotonically in our previous simulation
(\cite[Mitarai \& Nakanishi 2001; Mitarai \etal 2002]{MN01,MHN02}). 

For $\nu_0(0) \lesssim 0.10$, the higher density
flow shows lower flowing speed in the case of $\theta=16^\circ$,  
which results in the decrease of the flux $Q_0$ as 
$\bar \rho_0$ increase for $\bar \rho_0\lesssim 0.2$ 
(figure \ref{steadyphase005}(b)).
For higher density ($\nu_0(0)=0.15$ in figure \ref{steady}),
the flowing speed gets faster with $\bar \rho_0$, which causes the 
increase of the flux $Q_0$ for higher $\bar \rho_0$.
As a result, $Q_0$ has a minimum at a finite density.

In the case of  $\theta\le 14^\circ$, 
the velocity continuously decrease as the density becomes lower,
and $Q_0$ becomes a increasing function of $\bar \rho_0$.

\subsubsection{The case of $\phi=0.10$}
\begin{figure}
\begin{minipage}{0.5\textwidth}
\includegraphics[width=0.8\textwidth]{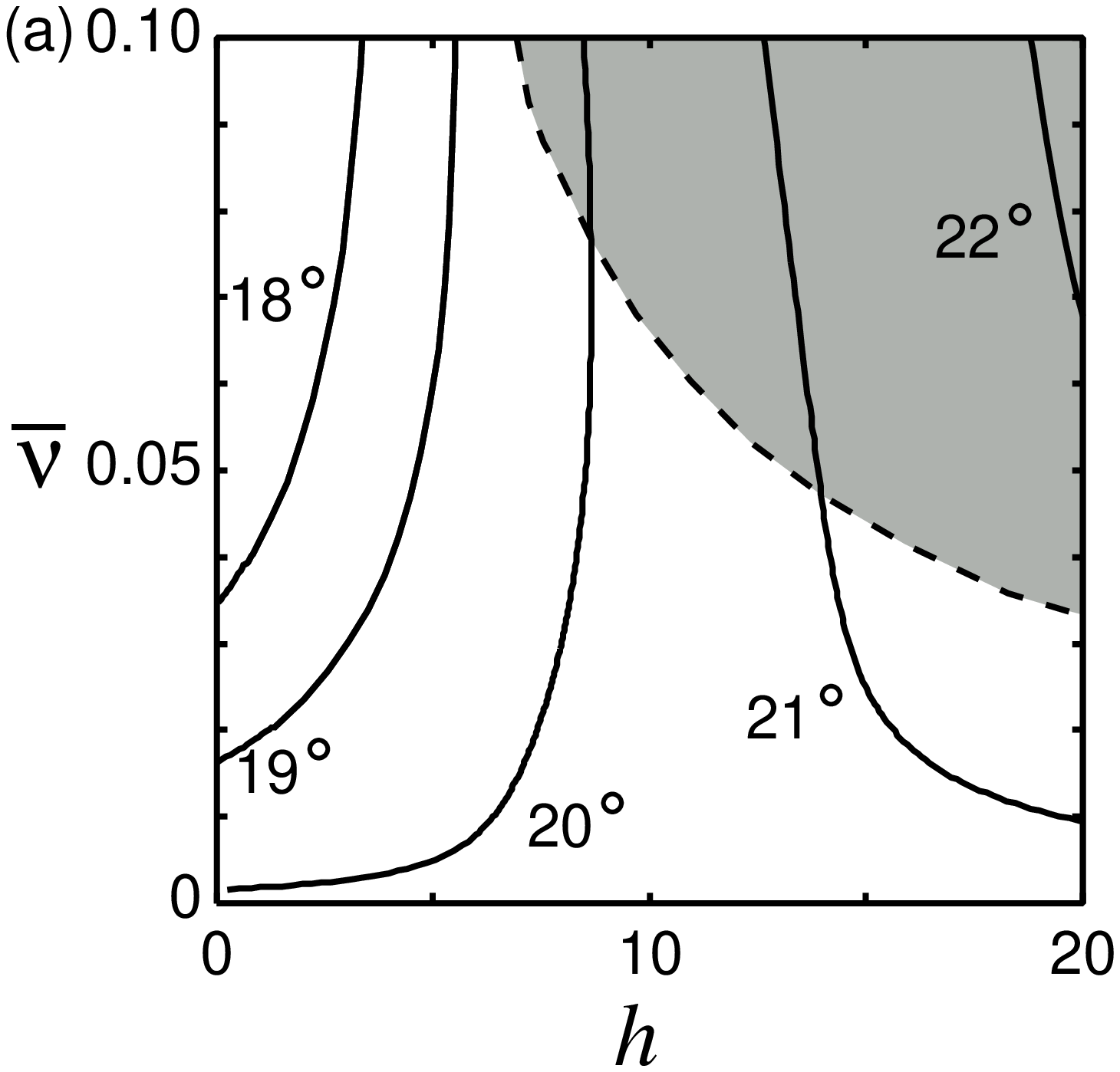}
\end{minipage}
\begin{minipage}{0.5\textwidth}
\includegraphics[angle=-90,width=\textwidth]{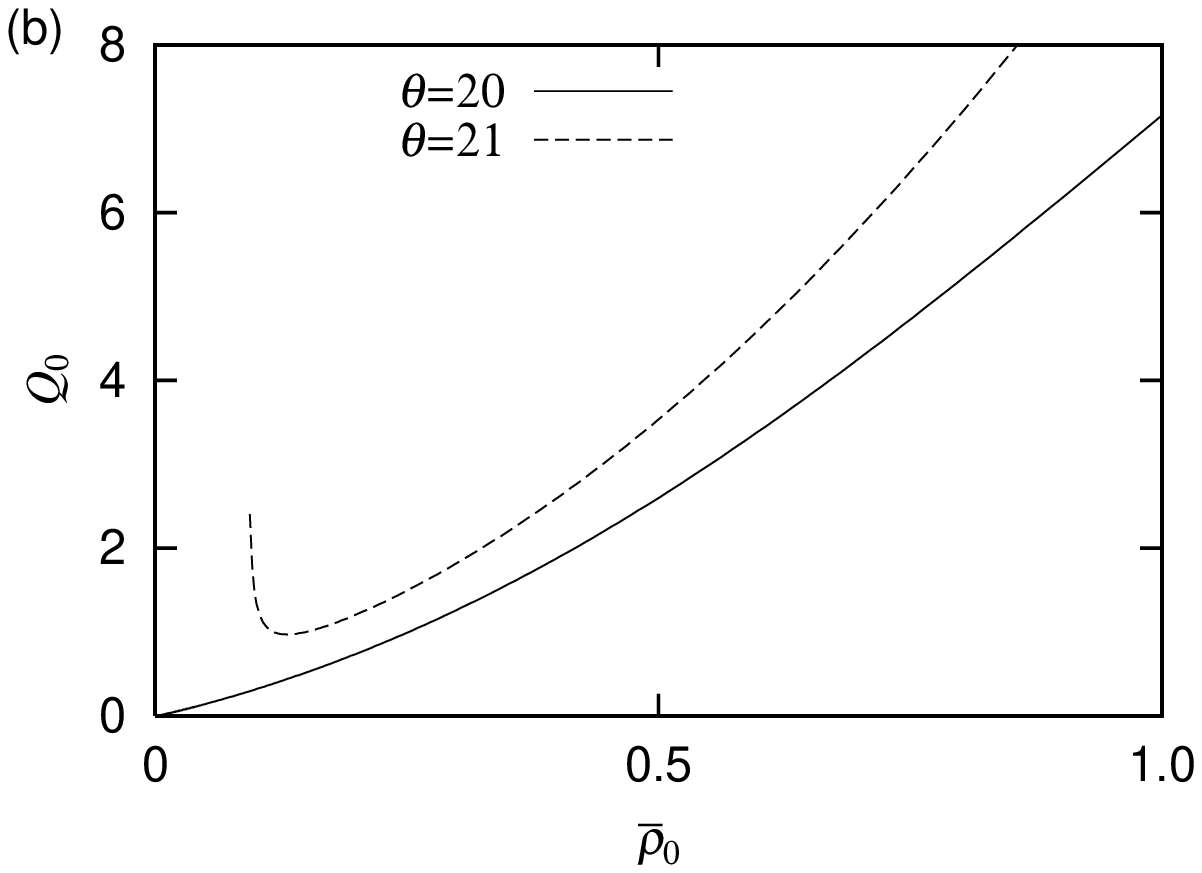}
\end{minipage}
\caption{(a)The contour lines of $\theta$
in the $(h,\bar \nu)$ plane for $\phi=0.10$.
The region of non-monotonic density profiles
is shown by a grey region.
(b)Relation between $Q_0$ vs $\bar \rho_0$ for $\phi=0.10$.
The solid line and the dashed line
are for $\theta=20^\circ$ and $\theta=21^\circ$, respectively.
}
\label{steadyphase010}
\end{figure}

\begin{figure}
\includegraphics[angle=-90,width=0.32\textwidth]{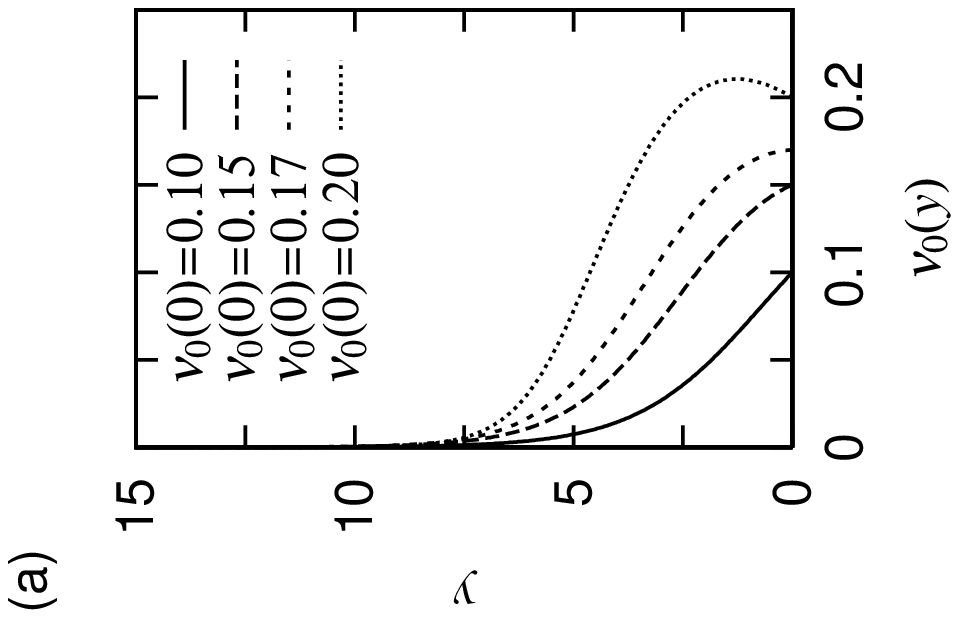}
\includegraphics[angle=-90,width=0.32\textwidth]{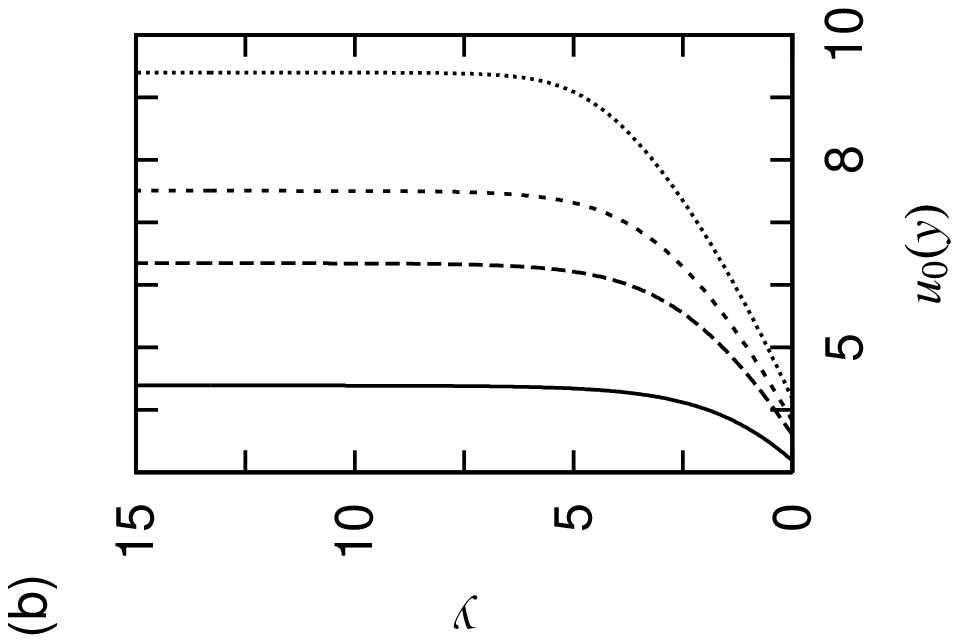}
\includegraphics[angle=-90,width=0.32\textwidth]{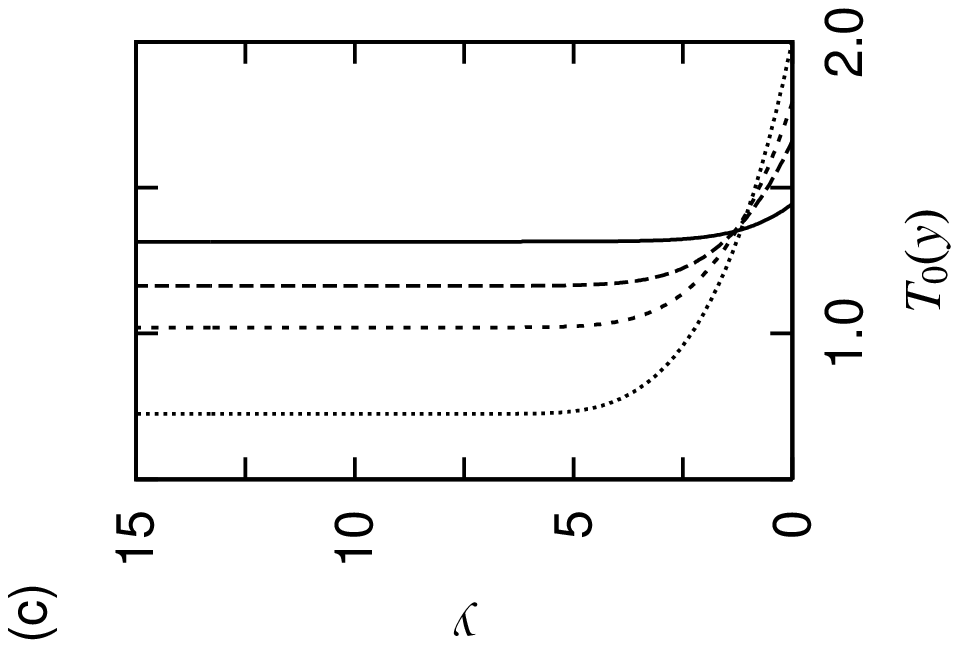}
\caption{The steady solutions for $\phi =0.10$ and $\theta=20^{\circ}$.
The density profiles (a), the velocity profiles (b),
and the temperature profiles (c).
Different line types correspond to the solutions with different
densities; $\nu_0(0)=0.10$ (solid lines),
$0.15$ (long-dashed lines)
$0.17$ (short-dashed lines), 
and $0.20$ (dotted lines). 
}
\label{steady01}
\end{figure}

The slope is rougher than the previous case, and
the steady solution 
exists for $12^{\circ}\lesssim \theta \lesssim 25^{\circ}$.
The contour lines for $\theta$ 
in the $(h,\bar \nu)$ plane are shown in figure \ref{steadyphase010}(a).
As in the case of $\phi=0.05$,
$h$ goes to zero as density becomes lower for smaller $\theta$,
while $h$ diverges for smaller density when $\theta$ is large enough.
Figure \ref{steadyphase010}(b) shows 
$\bar \rho_0$ dependence of $Q_0$,
which is a monotonically increasing function for $\theta = 20^\circ$
and has a minimum for $\theta= 21^\circ$.
The typical solutions are shown for 
$\theta=20^\circ$ in figure \ref{steady01}.
For large enough $h$ and $\bar \nu$,
the maximum density appears at a finite distance from the floor.
The region of the non-monotonic density profile 
is shown by a grey region in figure \ref{steadyphase010}(a).


\section{Linear Stability Analysis; Density Wave Formation}\label{lstaba}
\subsection{Normal mode analysis}
We restrict our stability analysis to the perturbation
uniform along the $z$ direction,
because we are interested in the 
instability along the flow direction.
The flow is perturbed around the steady solution 
as 
\begin{eqnarray}
\nu(x,y,t)&=&\nu_0(y)+\nu_1(x,y,t), \label{eq:per1}\\
\boldsymbol u(x,y,t)&=&(u_0(y),0,0)+(u_1(x,y,t),v_1(x,y,t)),0),
\label{eq:per2}\\
T(x,y,t)&=&T_0(y)+T_1(x,y,t).\label{eq:per3}
\end{eqnarray}
The governing equations and the boundary conditions 
are linearized with respect to
the deviations $\nu_1$, $u_1$, $v_1$, and $T_1$,
but the resulting expressions are rather lengthy and
given in Appendix A.

Now we look for the normal modes for the density,
the velocity, and the temperature perturbations
of the form 
\begin{equation}
(\nu_1,u_1,v_1,T_1)=\mbox{Re}\left[
\left(
\hat \nu(y),\hat u(y),\hat v(y),\hat T(y)
\right)\exp(\alpha t+ikx)
\right].
\label{eq:lin}
\end{equation}
The flow is linearly unstable
if $\mbox{Re}(\alpha)>0$.

As for the boundary condition at the free surface, 
the asymptotic behavior of the perturbations at large $y$
are used. When $y$ is much larger than
the decay length of the density and thus $\nu_0(y)$
is very small, the density perturbation should also decay 
($\hat \nu \propto \nu_0$), and $\hat u$, $\hat v$, and $\hat T$ 
decay proportional to $\exp(-ky)$ 
(\cite[Forterre \& Pouliquen 2002]{FP02}), therefore,
we imposed the boundary condition that
\begin{equation}
\hat u'(y)=-k\hat u(y),
\quad \hat v'(y)=-k \hat v(y),\quad
\hat T'(y)=-k\hat T(y),
\end{equation}
at $y=y_{\max}$; $y_{\max}$ is 
a large enough height where
$\nu_0(y_{\max})<10^{-9}$, 
in addition to the condition $|T'_0(y_{\max})|<10^{-7}$ 
discussed in \S \ref{secsteady}.

We solve the eigenvalue problems of
the linearized equations (\ref{eq:leqnu})-(\ref{eq:leqT})
for (\ref{eq:lin})
numerically using the Chebychev collocation method 
with the discretization in the $y$ direction 
(\cite[Gottilieb, Hussaini \& Orsag 1984]{spectral};
\cite[Canuto \etal 1988]{canuto};
\cite[Boyd 2001]{boyd};
\cite[Foterre \& Pouliquen 2002]{FP02}).
It is known that the straightforward discretization of space 
requires two extra boundary conditions 
(\cite[Malik 1990]{M90};
\cite[Foterre \& Pouliquen 2002]{FP02}), for which
we use the momentum balance condition in the $y$ direction at $y=0$
and the decay condition for the density perturbation, i.e., 
$\hat \nu'(y)=-(\cos\theta/T_0(y))\hat \nu(y)$ at $y=y_{\max}$.
In actual numerics to solve 
the generalized eigenvalue problem
in the form $A\boldsymbol V=\alpha B\boldsymbol V$
for the complex eigenvalues $\alpha$ and 
the eigenvectors $\boldsymbol V$,
we used LAPACK version 3.0 (\cite[Anderson \etal 1999]{laug}).
The number of discretization $N_d$ is about $100$.

The numerically obtained eigenmodes
contains unphysical modes, 
called spurious modes,
due to the discretization
(\cite[Mayer \& Powell 1992]{MP92};
\cite[Boyd 2001]{boyd};
\cite[Forterre \& Pouliquen 2002]{FP02}).
For the spurious modes, it is known that 
the Chebychev coefficients of higher wave number
are large, and
the eigenvalues are sensitive to small change of $N_d$. 
We determine the eigenmodes as physical ones
by checking that their 
Chebychev coefficients for higher wave number
are small and their eigenvalues 
varies little upon changing $N_d$: 
We confirmed that,
for these modes,
the highest ten coefficients are less than $10^{-7}$
when the eigenvectors are normalized so that 
the sum of the absolute values of the real part 
and the imaginary part of the largest component becomes one,
and the variation of the eigenvalues
by the small change of $N_d$ are less than $10^{-7}$.

\subsection{Stability diagram, dispersion relations, 
and eigenfunctions}\label{disp005}
We present the results of the linear stability analysis 
in the cases of $\phi=0.05$ and $\phi=0.10$, 
for which the steady solutions are shown in \S \ref{steadysol}.

\subsubsection{The case of $\phi=0.05$}
\begin{figure}
\begin{center}
\includegraphics[width=.4\textwidth]{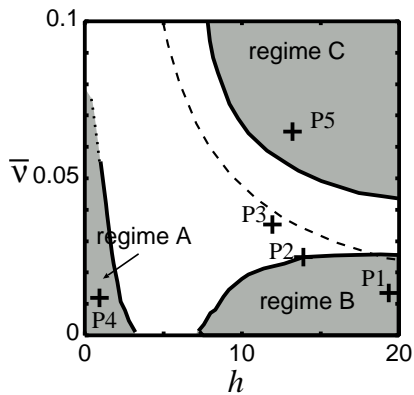}
\end{center}
\caption{The stability diagram for $h$ vs $\bar \nu$ for $\phi =0.05$,
where the unstable (stable) regimes are shown by grey (white) regions.
We find three unstable regimes A, B, and C.
The dashed line shows the boundary between 
the region of the non-monotonic density
profiles and that of the monotonic density profiles (see also figure
 \ref{steadyphase005}).
The dispersion relations at the points P1$\sim$P5 are shown
in figures \ref{disp} $\sim$ \ref{disp-AC}.
The region of $h<1$ where the stability boundary 
is shown by a dotted line was not examined in detail 
because the flow with the
decay length less than the particle diameter is physically unacceptable. 
}
\label{phase005}
\end{figure}
\begin{figure}
\includegraphics[angle=-90,width=0.32\textwidth]{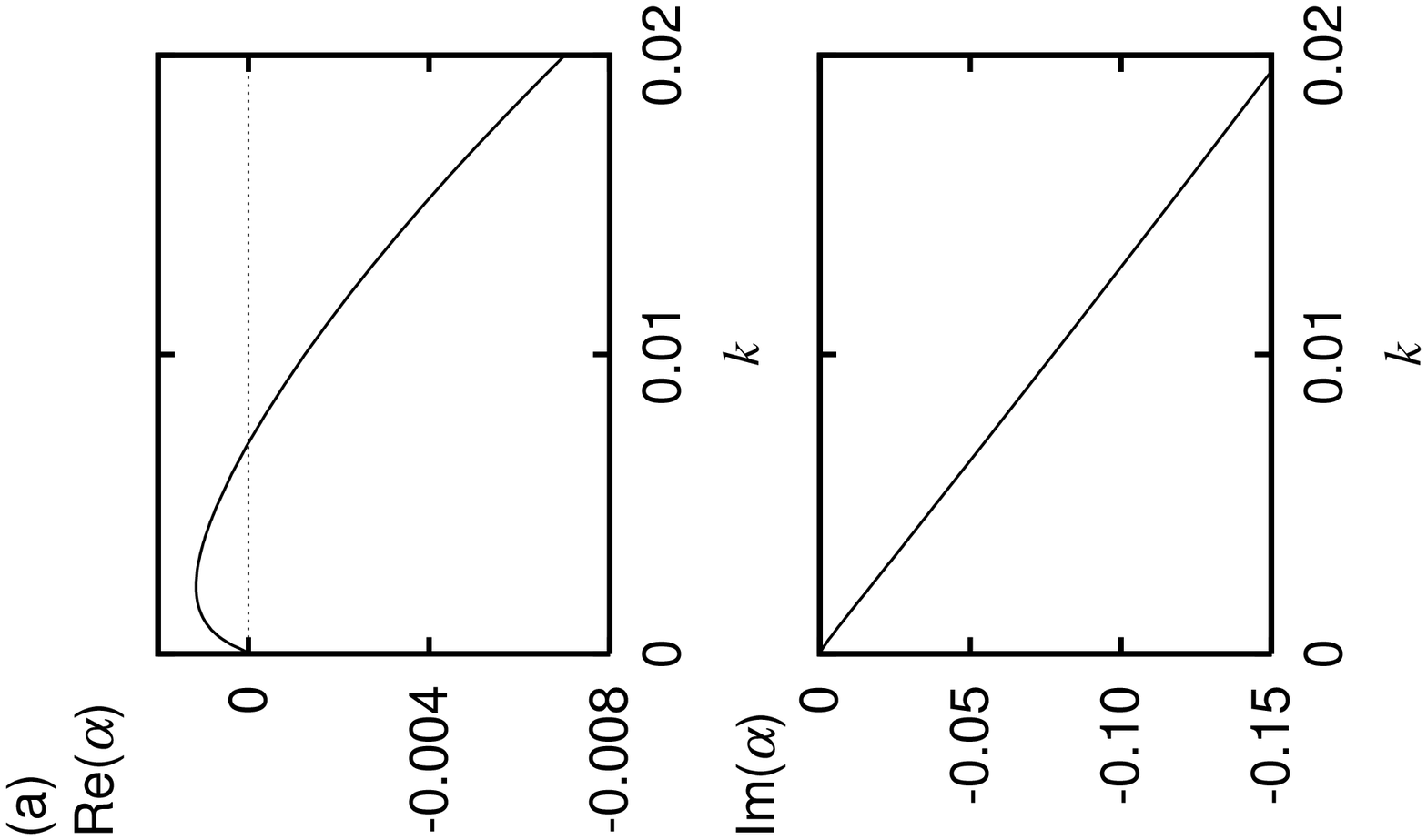}
\includegraphics[angle=-90,width=0.32\textwidth]{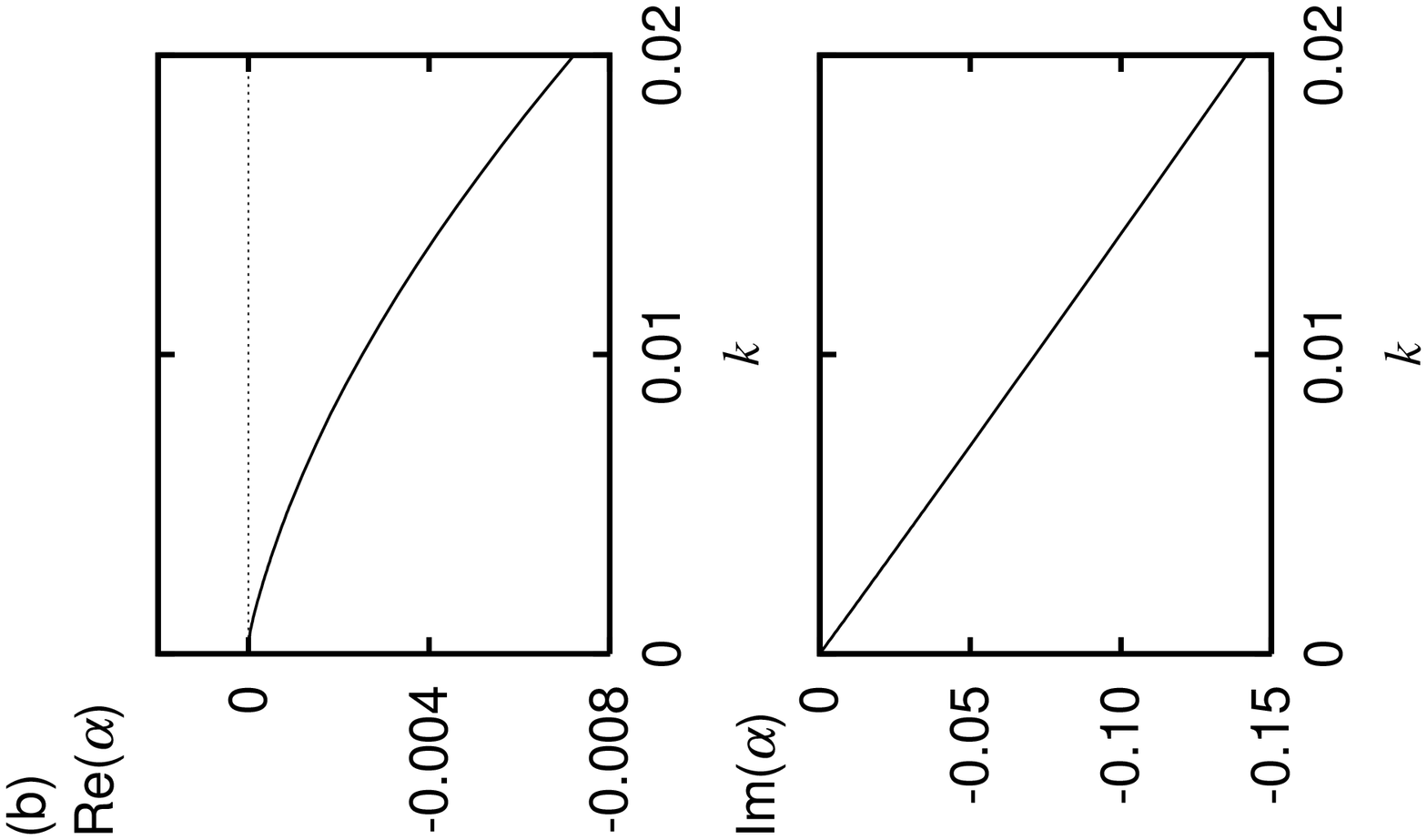}
\includegraphics[angle=-90,width=0.32\textwidth]{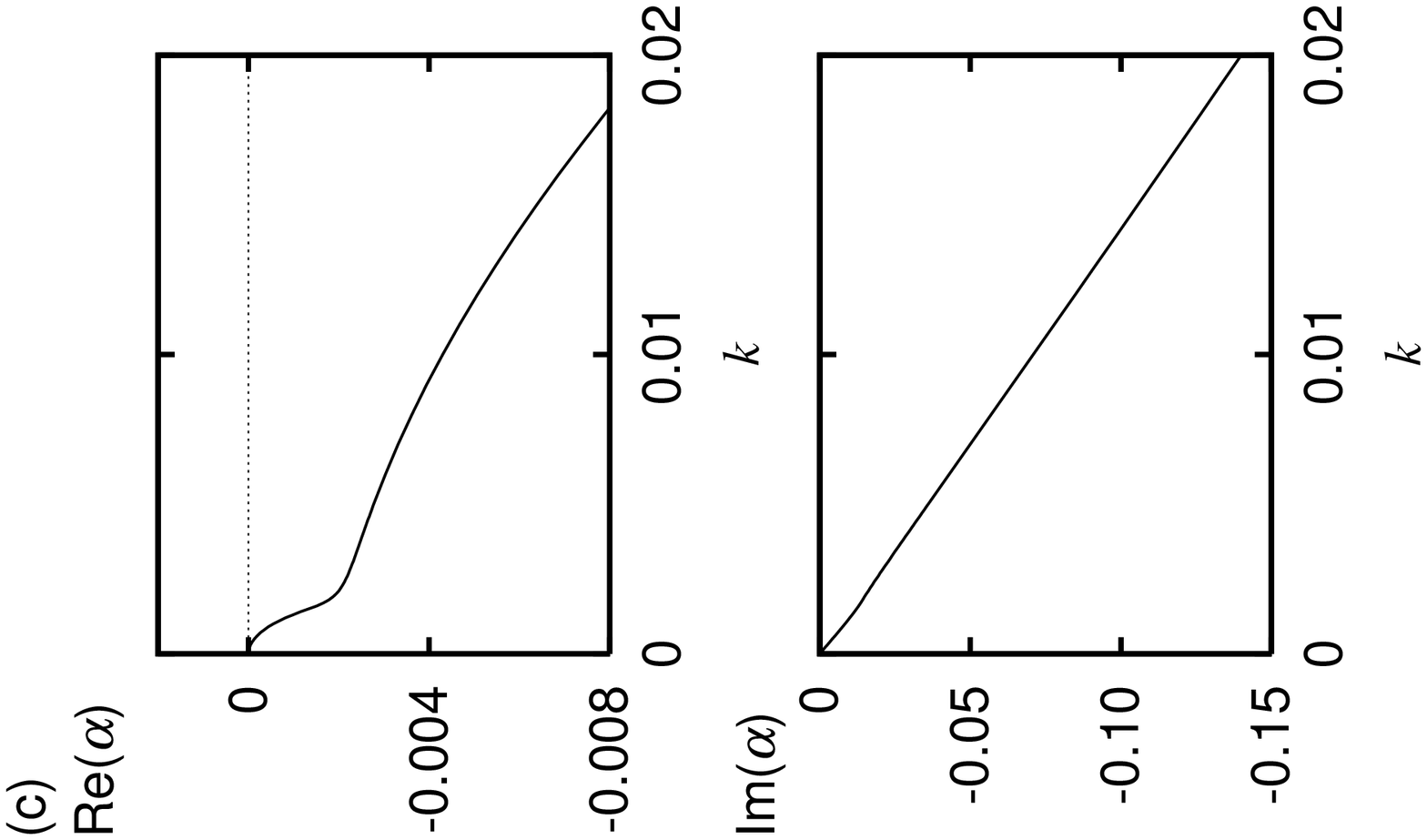}
\caption{The dispersion relations of the least stable mode
for the steady solutions with $\phi =0.05$ and 
$\theta=16^{\circ}$ for $\nu_0(0)=0.05$ (a),
$0.08$ (b), and $0.10$ (c),
which correspond
to the points P1, P2, and P3 in figure \ref{phase005}, respectively.}
\label{disp}
\end{figure}
\begin{figure}
\begin{center}
\includegraphics[angle=-90,width=0.4\textwidth]{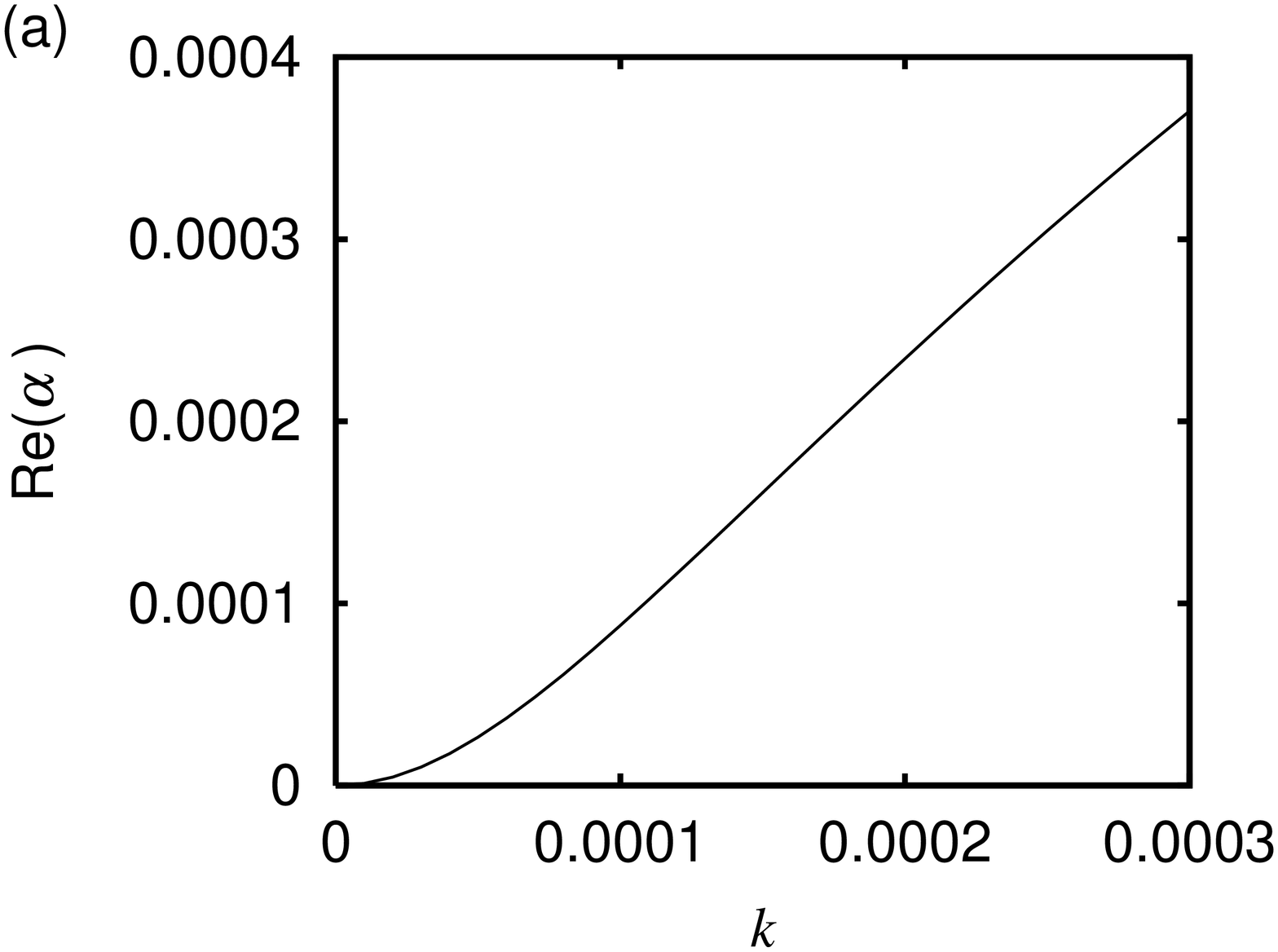}
\includegraphics[angle=-90,width=0.4\textwidth]{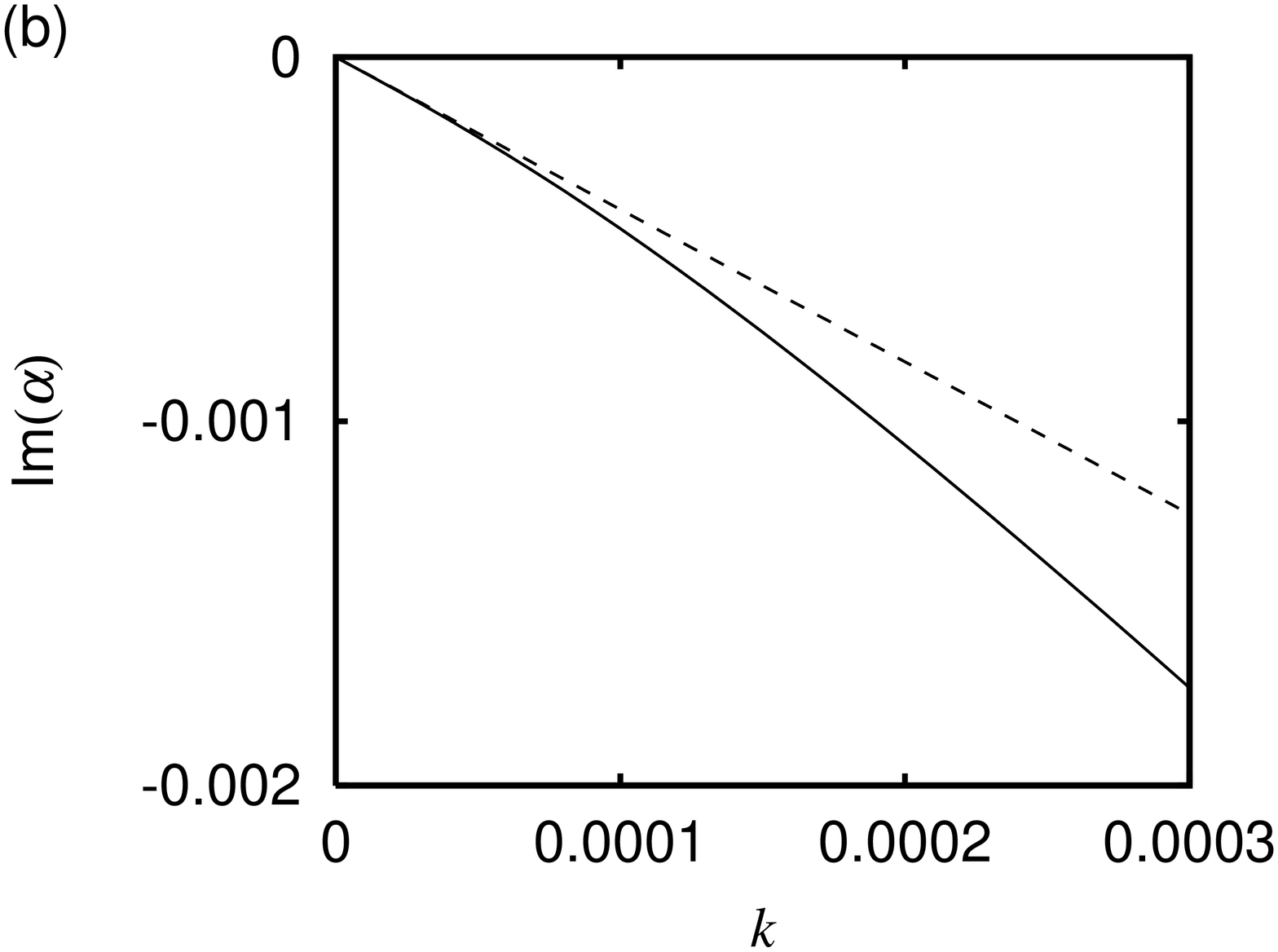}
\end{center}
\caption{The magnification of the dispersion relations near $k=0$
of the unstable stable mode for the steady solutions with $\phi =0.05$ 
and $\theta=16^{\circ}$ for $\nu_0(0)=0.05$ 
at P1 in the regime B. 
Re[$\alpha$] grows quadratically in $k$, while the 
slope of Im[$\alpha$] is given by $-\mbox{d}Q_0/\mbox{d}\bar \rho_0$ in the
long-wavelength limit,  which is shown by a dashed line.
}
\label{dispmag}
\end{figure}
\begin{figure}
\begin{center}
\includegraphics[angle=-90,width=0.4\textwidth]{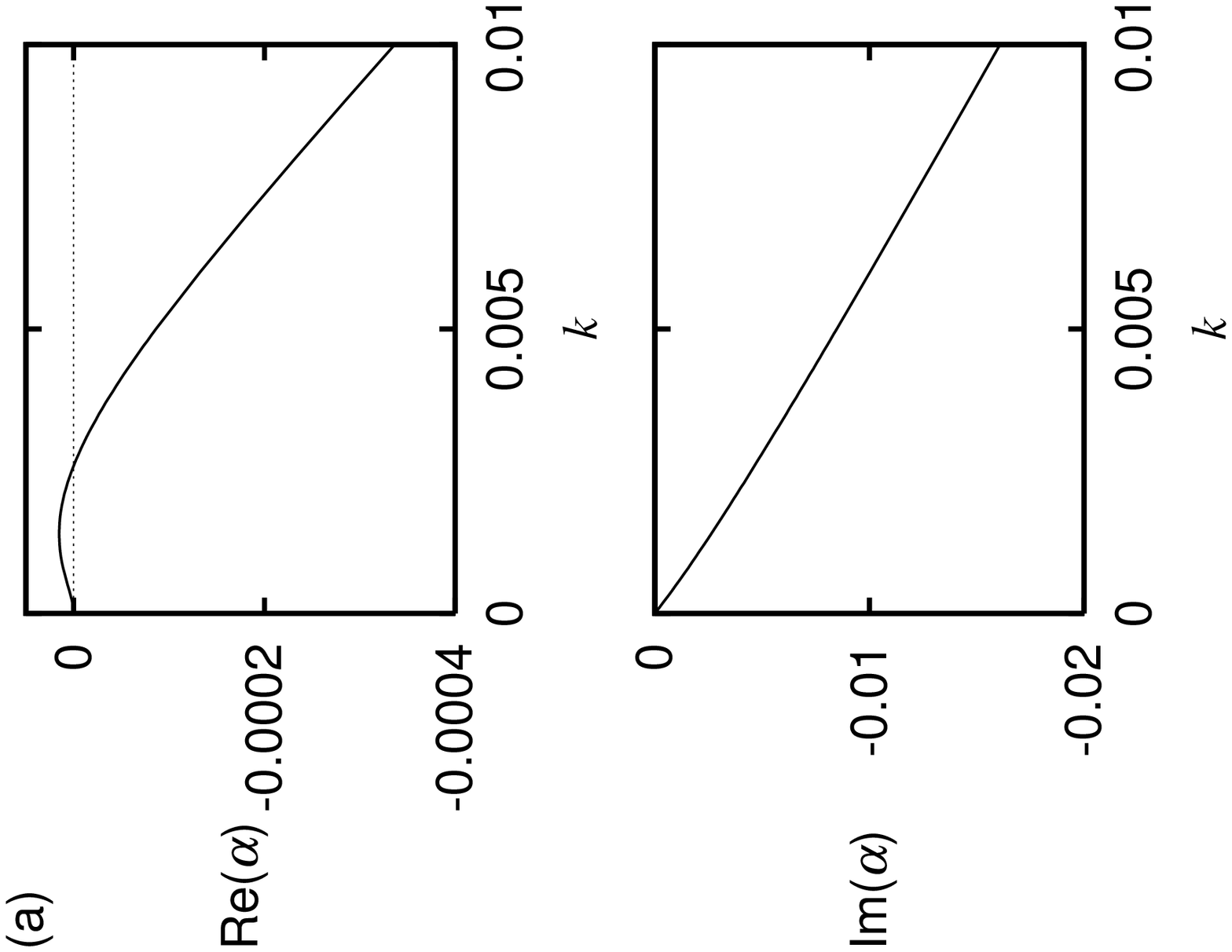}
\includegraphics[angle=-90,width=0.4\textwidth]{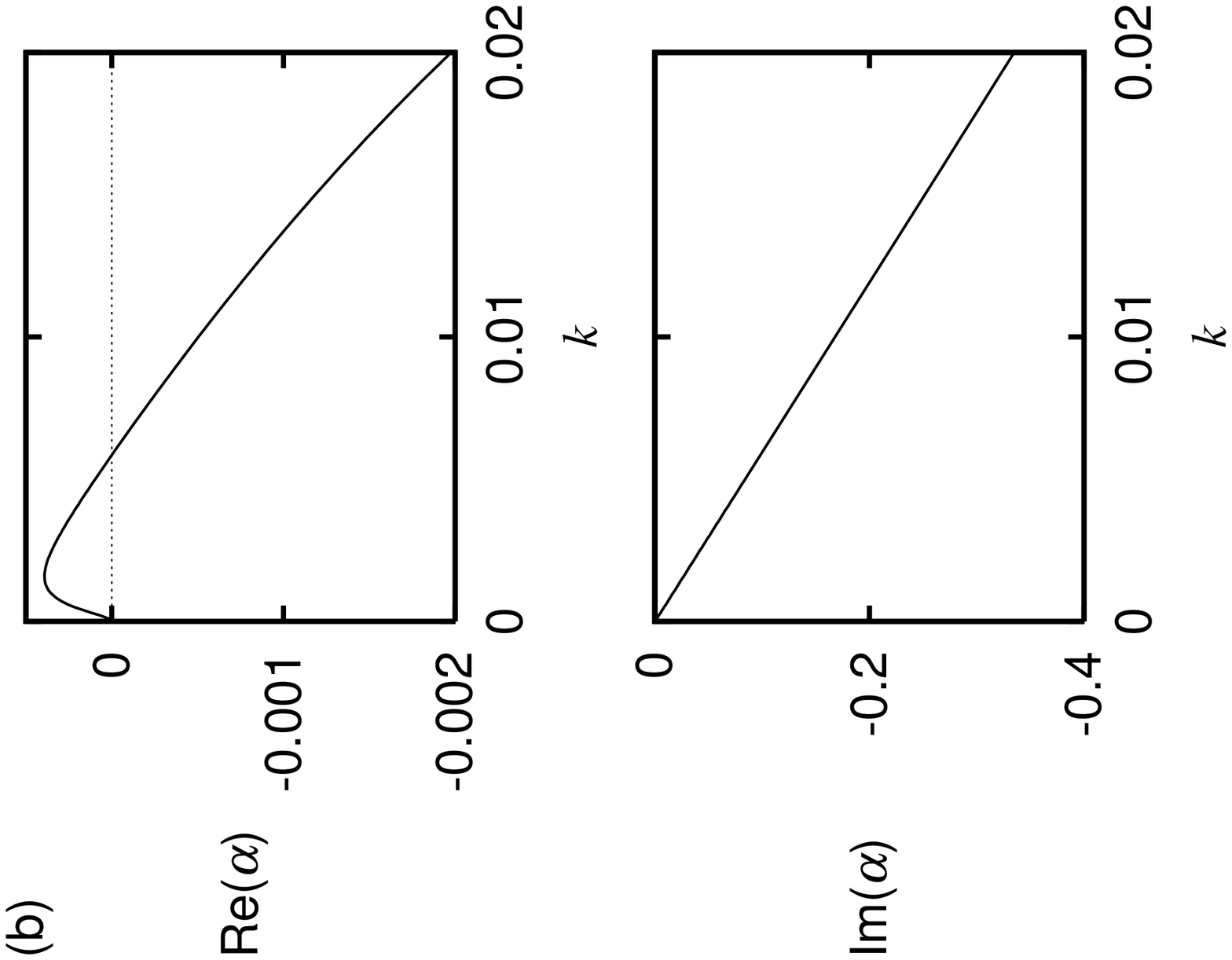}
\end{center}
\caption{The dispersion relations of the unstable mode
for $\phi =0.05$ with
(a) $\theta=14^{\circ}$ and $\nu_0(0)=0.05$ (P4; in regime A)
and (b) $\theta=17^{\circ}$ and $\nu_0(0)=0.10$ (P5; in regime C).}
\label{disp-AC}
\end{figure}

The stability diagram is shown in figure \ref{phase005}
in the parameter space of $h$ vs $\bar \nu$.
The unstable (stable) regimes are shown by grey (white) 
regions,
and the boundary between the region of the monotonic 
density profiles
and that of the non-monotonic density profiles 
is shown by a dashed line.
Within the investigated region, we find 
three unstable regimes;
the regime A at small $h$ and small $\bar \nu$ region,
the regime B at large $h$ and small $\bar \nu$ region,
and the regime C at large $h$ and large $\bar \nu$ region.
When we decrease the density 
with a constant inclination angle
(along a contour in figure \ref{steadyphase005}),
eventually we will encounter either the regime A or the regime B, 
namely, the flow with low enough density is always unstable.
The two regimes are different in the steady flow behavior
as we have seen already: 
in the regime A at the small $h$ side, 
the flow becomes slower as the density becomes smaller,
while the flow in the regime B at
the large $h$ side flows down faster for the smaller density.
On the other hand, 
the denser flow can be unstable
in the regime C, 
which lies within the region of 
the non-monotonic density profiles.
In this regime, the denser flow goes faster as in the regime A.

The dispersion relations of the least stable modes
$\alpha=\alpha(k)$ are shown in figure \ref{disp} 
for $\theta=16^{\circ}$ and 
$\nu_0(0)=0.05$ (a), $0.08$ (b),  and $0.10$ (c),
which correspond to the points 
P1, P2, and P3, respectively, in the stability diagram of
figure \ref{phase005}; 
P1 lies within the unstable regime B.
In all cases, it is found that the least stable mode satisfies
$\alpha(0)=0$. 
The growth rate $\mbox{Re}(\alpha)$ is positive for $\nu_0(0)=0.05$ 
for $0<k\le k_c$ with $k_c\approx 0.007$ (figure \ref{disp}(a)).
The magnification around $k=0$ in figure \ref{dispmag}
shows that $\mbox{Re}(\alpha)$ grows quadratically in $k$ for small $k$.
As $\nu_0(0)$ is increased, 
the maximum value of $\mbox{Re}[\alpha]$ 
decreases and becomes negative for all $k$ (figure \ref{disp}(b) and
(c)). 

The dispersion relations for the unstable modes
at P4 (in the regime A) 
and at P5 (in the regime C) are shown in figure \ref{disp-AC}
for $(\theta,\nu_0(0))=(14^\circ, 0.05)$ (a)
and $(17^\circ, 0.10)$ (b), respectively.
The instability occurs for the long-wavelength perturbation,
and both of the dispersion relations show that
$\alpha(0)=0$ and 
the growth rate quadratic in $k$ for small $k$, i.e.,
$\mbox{Re}[\alpha(k)]\propto k^2$;
these features are the same as those in the unstable regime B.

\begin{figure}
\begin{center}
\includegraphics[width=.522\textwidth]{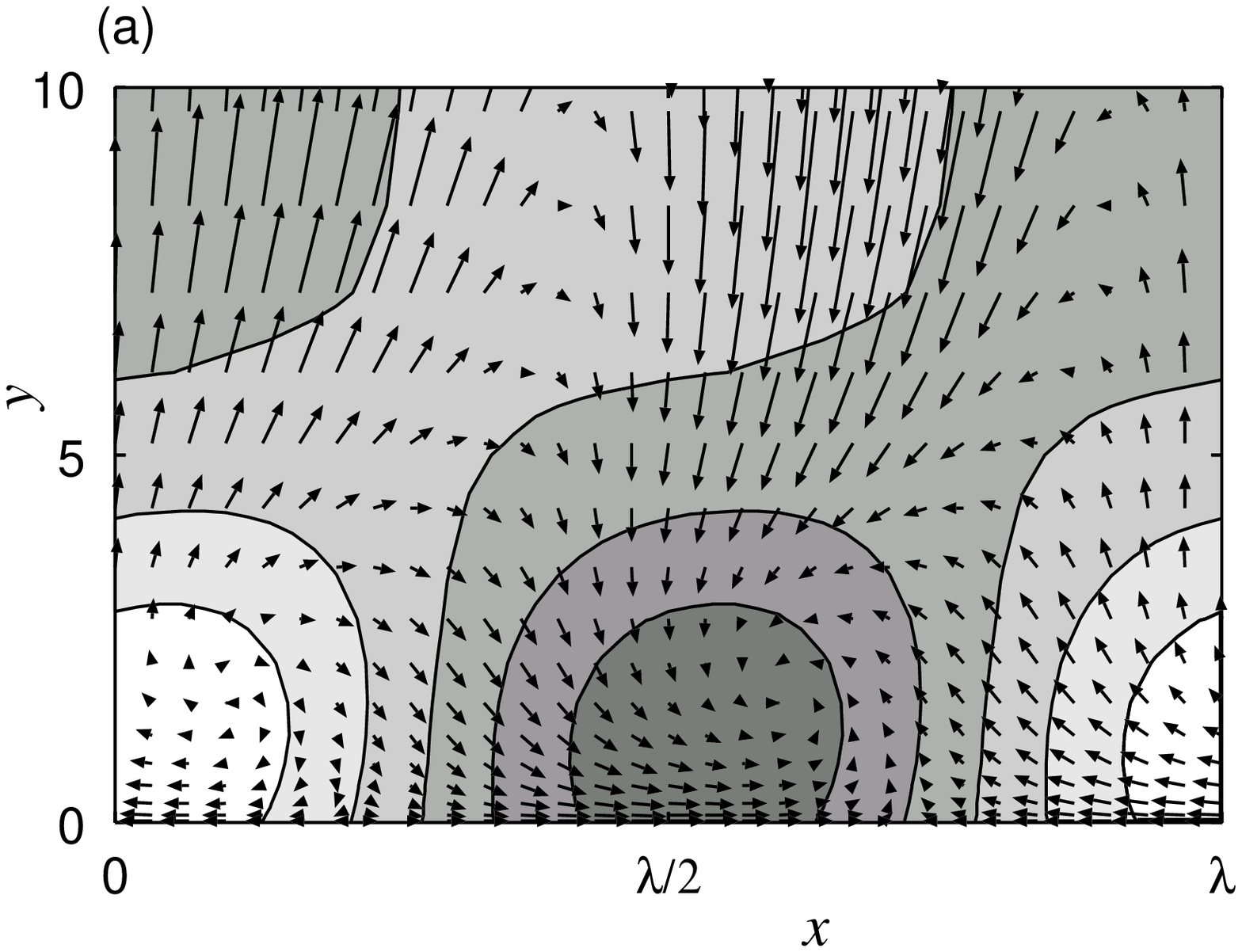}
\includegraphics[width=.47\textwidth]{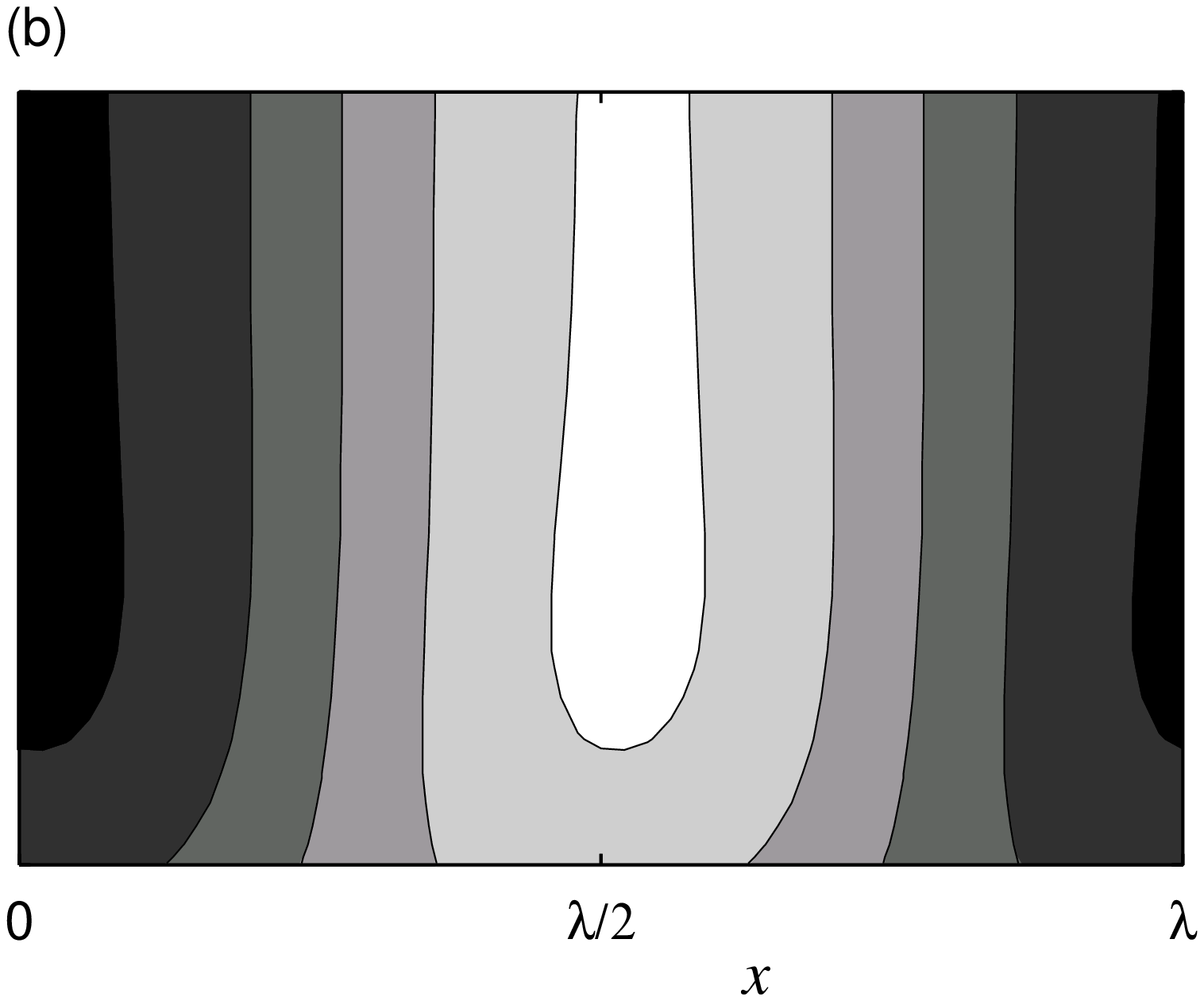}
\end{center}
\caption{The eigenfunctions of the least stable mode 
for $\phi=0.05$, $\theta=16^\circ$, and $\nu_0(0)=0.10$
(P3 in figure \ref{phase005}), 
whose wave number is $k=0.002$.
$\lambda$ is the wavelength of this eigenmode, $\lambda=2\pi/k$.
The contour of the density and the temperature eigenfunctions are
shown in (a) and (b), respectively, by grey scale, 
where the brighter (darker) region
corresponds to the larger positive (negative) value.
The arrows in (a) indicate the corresponding velocity eigenfunction.
It shows that the grains flow from the brighter region
into the darker density region, namely, the density perturbation decays.
Note that the temperature perturbation is negative at 
the region of the positive density perturbation.
}
\label{eigens}
\end{figure}
\begin{figure}
\begin{center}
\includegraphics[width=.522\textwidth]{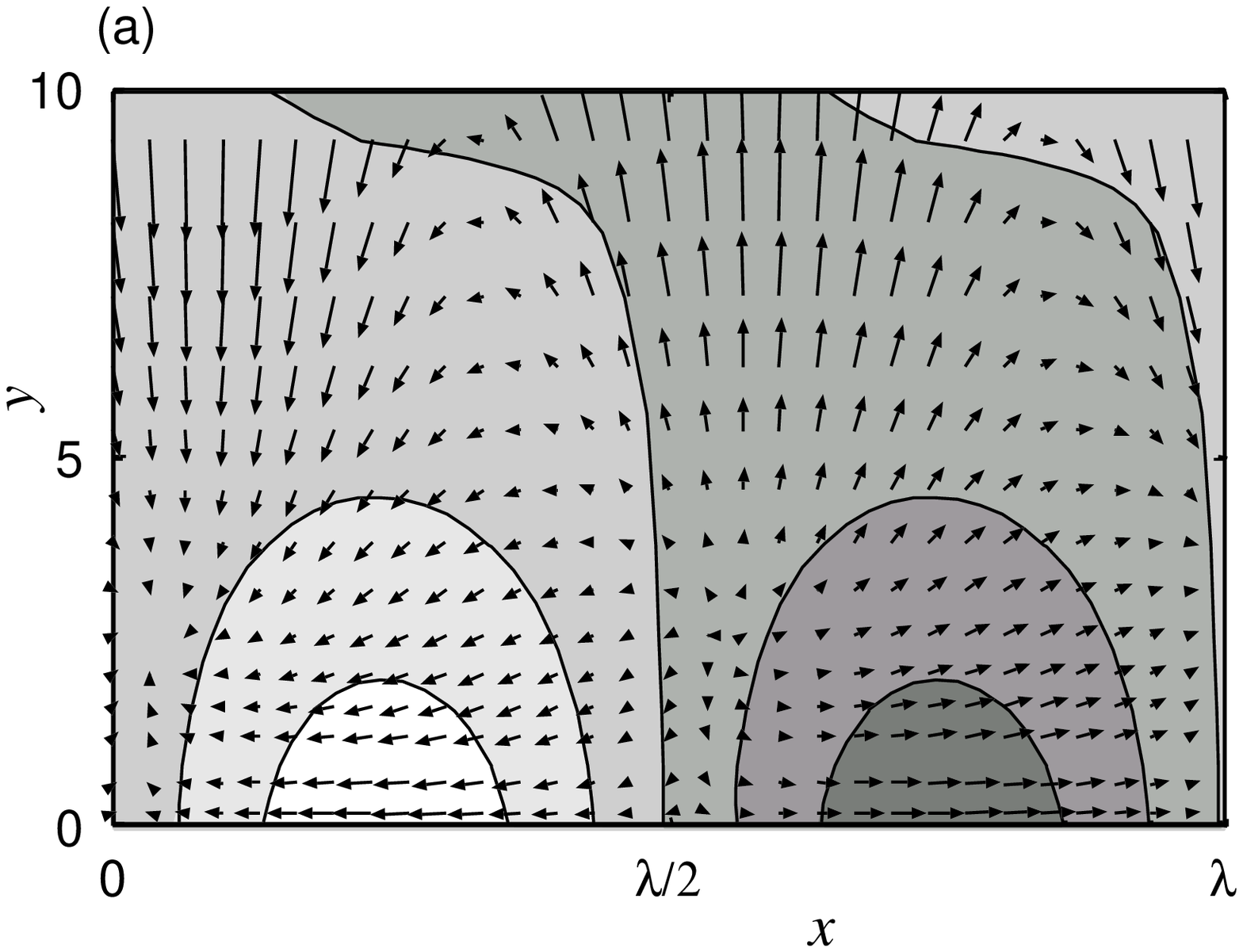}
\includegraphics[width=.47\textwidth]{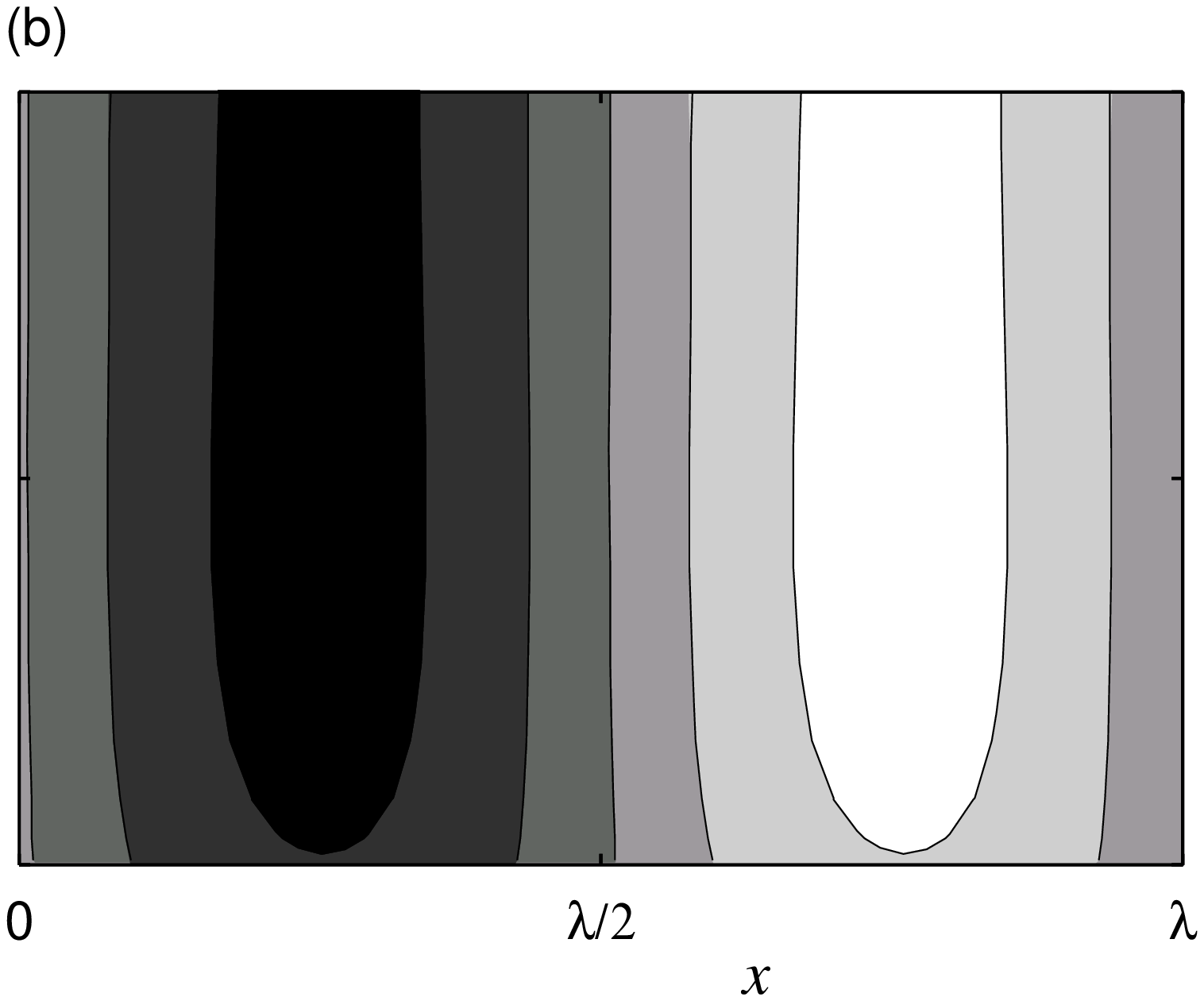}
\end{center}
\caption{The eigenfunctions of the unstable mode 
for $\phi=0.05$, $\theta=16^\circ$, and $\nu_0(0)=0.05$
(P1 in figure \ref{phase005}), 
whose wave number is $k=0.002$.
The contour of the density eigenfunction and 
the velocity eigenfunction in (a) 
show that the grains flow from the region of negative density perturbation
into that of positive perturbation, namely,
the perturbation is amplified and
results in the formation of the density wave.
The temperature perturbation in (b)
is negative at the region of positive density perturbation.
}
\label{eigenu}
\end{figure}
The least stable eigenmodes 
for $\theta=16^\circ$ at $k=0.002$
are shown for the two cases: 
the stable case of $\nu_0(0)=0.10$ 
(P3 in figure \ref{phase005})
in figure \ref{eigens} 
and the unstable case of $\nu_0(0)=0.05$ 
in the regime B
(P1 in figure \ref{phase005})
in figure \ref{eigenu},
over one wavelength $\lambda=2\pi/k$.
The contours in figures \ref{eigens}(a) and \ref{eigenu}(a)
show the density eigenfunctions;
the brighter (darker) regions indicate the positive 
(negative) regions, and the arrows
represent the corresponding velocity eigenfunctions.
In the both figures near $y=0$, we see that 
the velocity perturbations 
point to the positive (negative) $x$ direction
in the regions of negative (positive)
density perturbation.
The contours for the corresponding temperature
eigenfunctions plotted in 
figures \ref{eigens}(b) and \ref{eigenu}(b)
show that the regions where the density perturbation is negative 
(darker regions in figures \ref{eigens}(a) and \ref{eigenu}(a))
roughly correspond 
to the positive temperature perturbations
(brighter regions in figures \ref{eigens}(b) and \ref{eigenu}(b)).

The difference between the stable mode (figure \ref{eigens})
and the unstable mode (figure \ref{eigenu})
is seen if we focus on the divergence of velocity perturbation.
In the case of the stable mode for $\nu_0(0)=0.10$ (figure \ref{eigens}(a)), 
the grains flow into the region where the density perturbation is negative
(see the region around $x\approx\lambda/2$ and $y\approx 2$), 
thus the density perturbation has negative feedback.
On the other hand, in the case of unstable mode for $\nu_0(0)=0.05$
(figure \ref{eigenu}(a)), 
the grains flow into the region where the density perturbation is positive
(see the region around $x\approx 0$ and $y\approx 3$);
As a result, the perturbation grows and 
eventually causes the nonlinear density wave.

The eigenfunctions of the unstable modes 
at P4 in the regime A
and at P5 in the regime C show some different characteristics 
from those at P1 in the regime B.
Reflecting the difference in the steady flow 
between the regimes A and C and the regime B,
the denser parts of the density eigenfunctions roughly correspond
to the part where the velocity fluctuation has
the positive component in the direction 
parallel to the mean flow.
For all three regimes, however, the unstable modes show that
the grains flow into 
the region of the positive density perturbation;
this suggests that the instability leads to 
the density wave in the regimes A and C as in the regime B.

\subsubsection{The case of $\phi =0.10$} 
\begin{figure}
\begin{center}
\includegraphics[width=.4\textwidth]{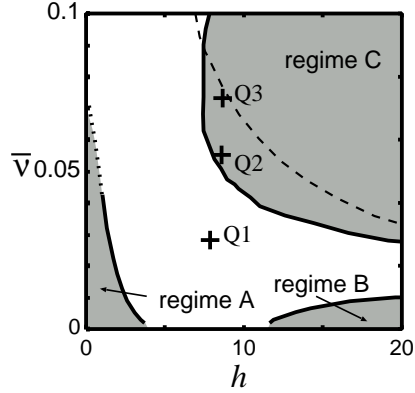}
\end{center}
\caption{The stability diagram for $h$ vs $\bar \nu$ with $\phi =0.10$,
where the unstable (stable) regimes are shown by grey (white) regions.
We find three unstable regimes A, B, and C.
The dashed line shows the boundary between the region of 
the non-monotonic density
profiles and that of the monotonic density profiles (see also figure
 \ref{steadyphase010}).
The dispersion relations at the points Q1$\sim$Q3 are shown
in figure \ref{disp01}.}
\label{phase010}
\end{figure}
\begin{figure}
\includegraphics[angle=-90,width=0.32\textwidth]{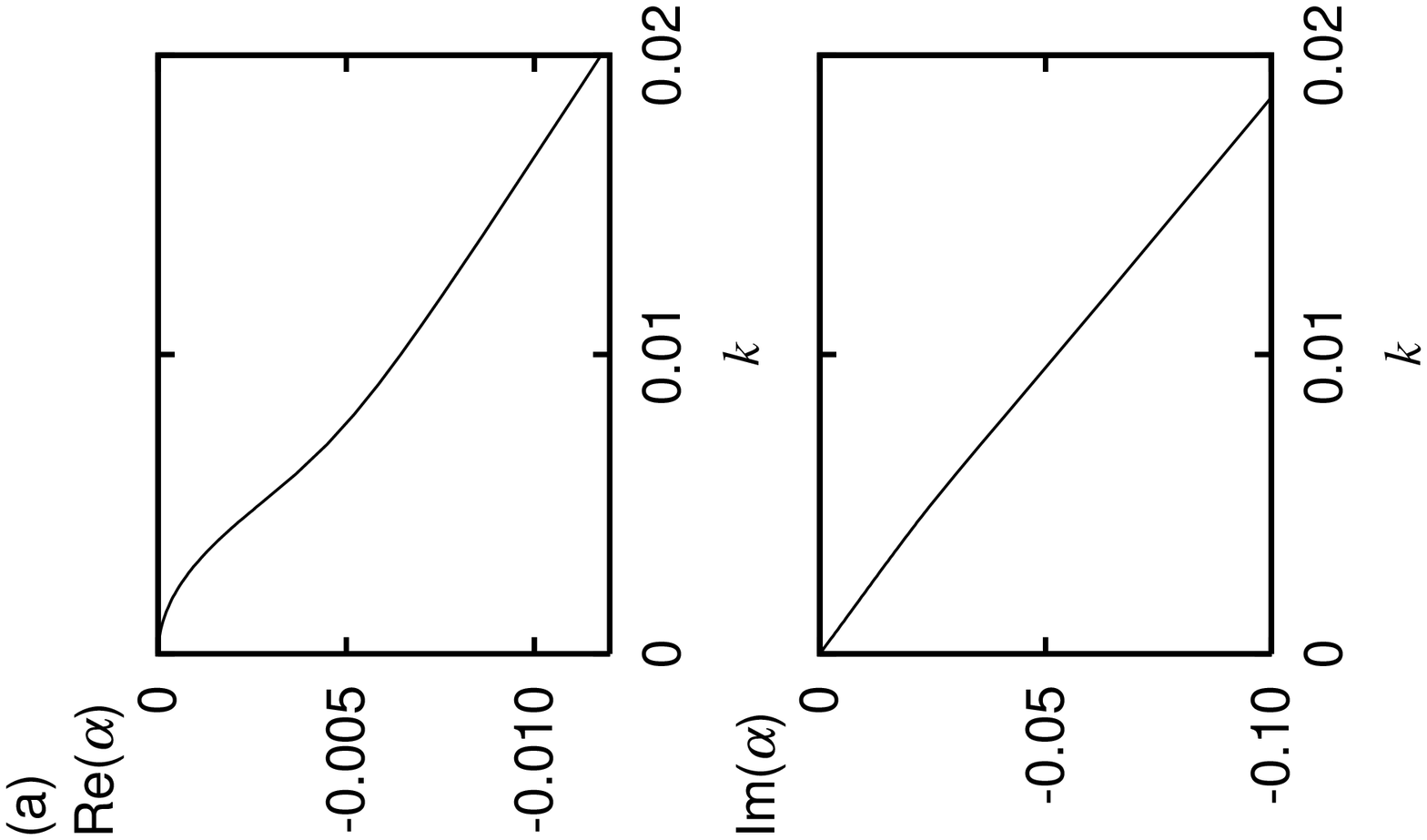}
\includegraphics[angle=-90,width=0.32\textwidth]{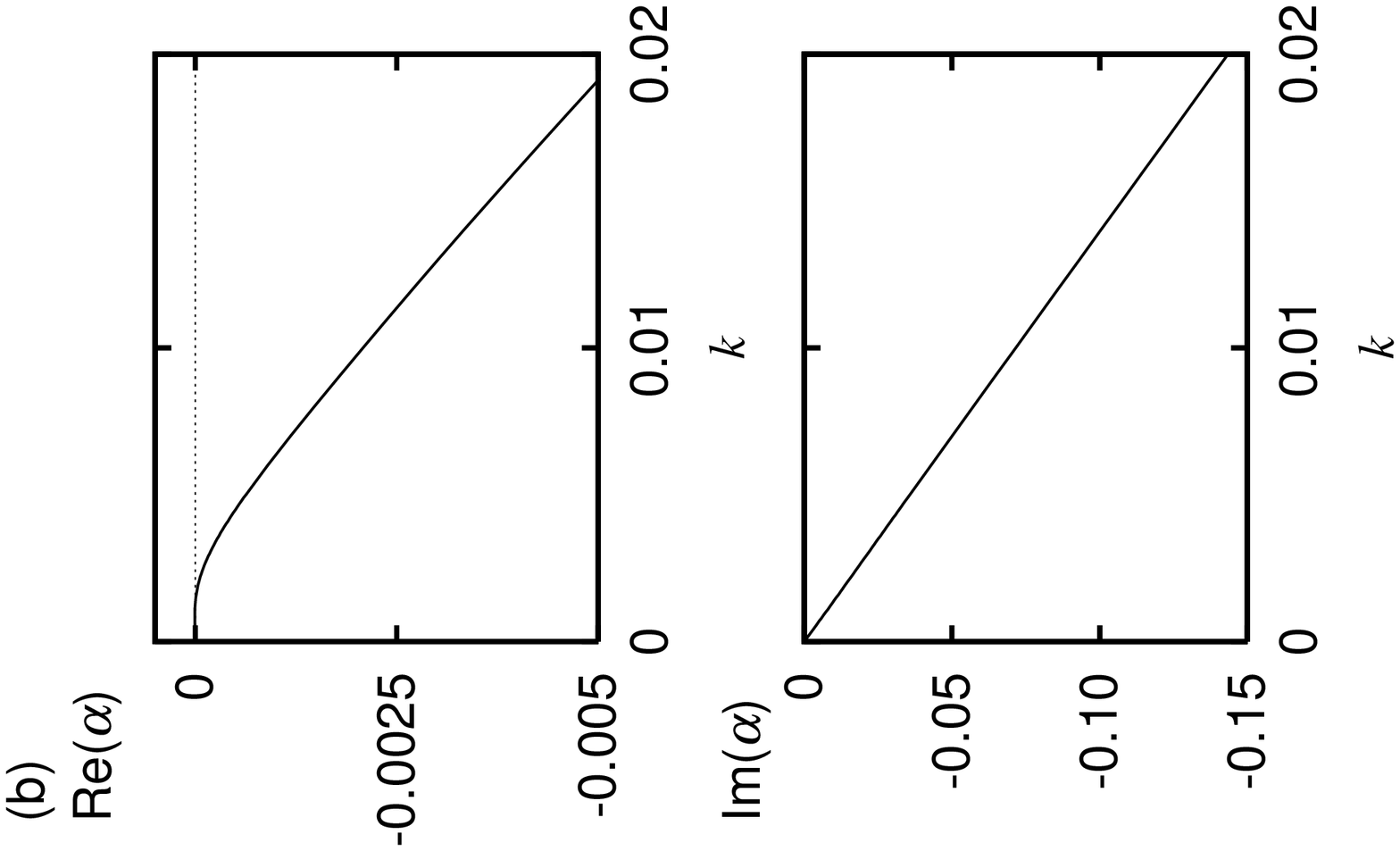}
\includegraphics[angle=-90,width=0.32\textwidth]{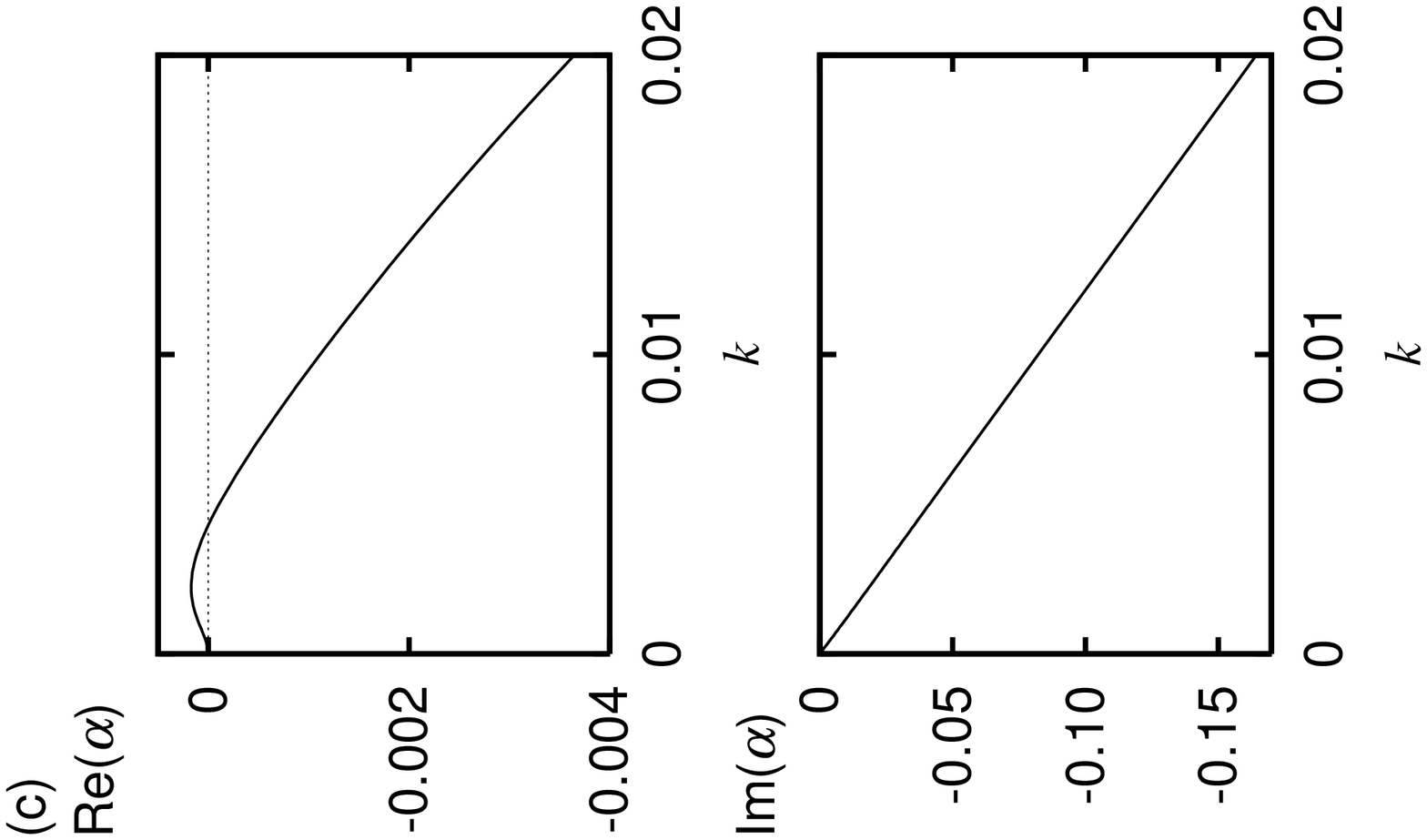}
\caption{
The dispersion relations of the least stable mode with $\phi =0.10$ 
and $\theta=20^{\circ}$ for $\nu_0(0)=0.10$ (a)
$0.15$ (b), and $0.17$ (c), which correspond
to the points Q1, Q2, and Q3 in figure \ref{phase010}, respectively.}
\label{disp01}
\end{figure}
The stability diagram for $\phi=0.10$
is shown in figure \ref{phase010}.
As in the case of $\phi=0.05$, there are 
three unstable regimes,
but the qualitative difference is that 
the unstable regime C at the large $h$ and large $\bar \nu$ region
contains a part of the region of the monotonic density profiles
as well as that of the non-monotonic density profiles.
The dispersion relations around the boundary of the regime C
are shown in figure \ref{disp01} 
for $\theta=20^\circ$ and
$\nu_0(0)=0.10$(a), $0.15$(b) and $0.17$(c),
which correspond 
to the points Q1, Q2, and Q3, respectively, 
in the stability diagram \ref{phase010}. 
It is seen that the instability occurs against 
the long-wavelength perturbation. 

\begin{figure}
\begin{center}
\includegraphics[width=0.522\textwidth]{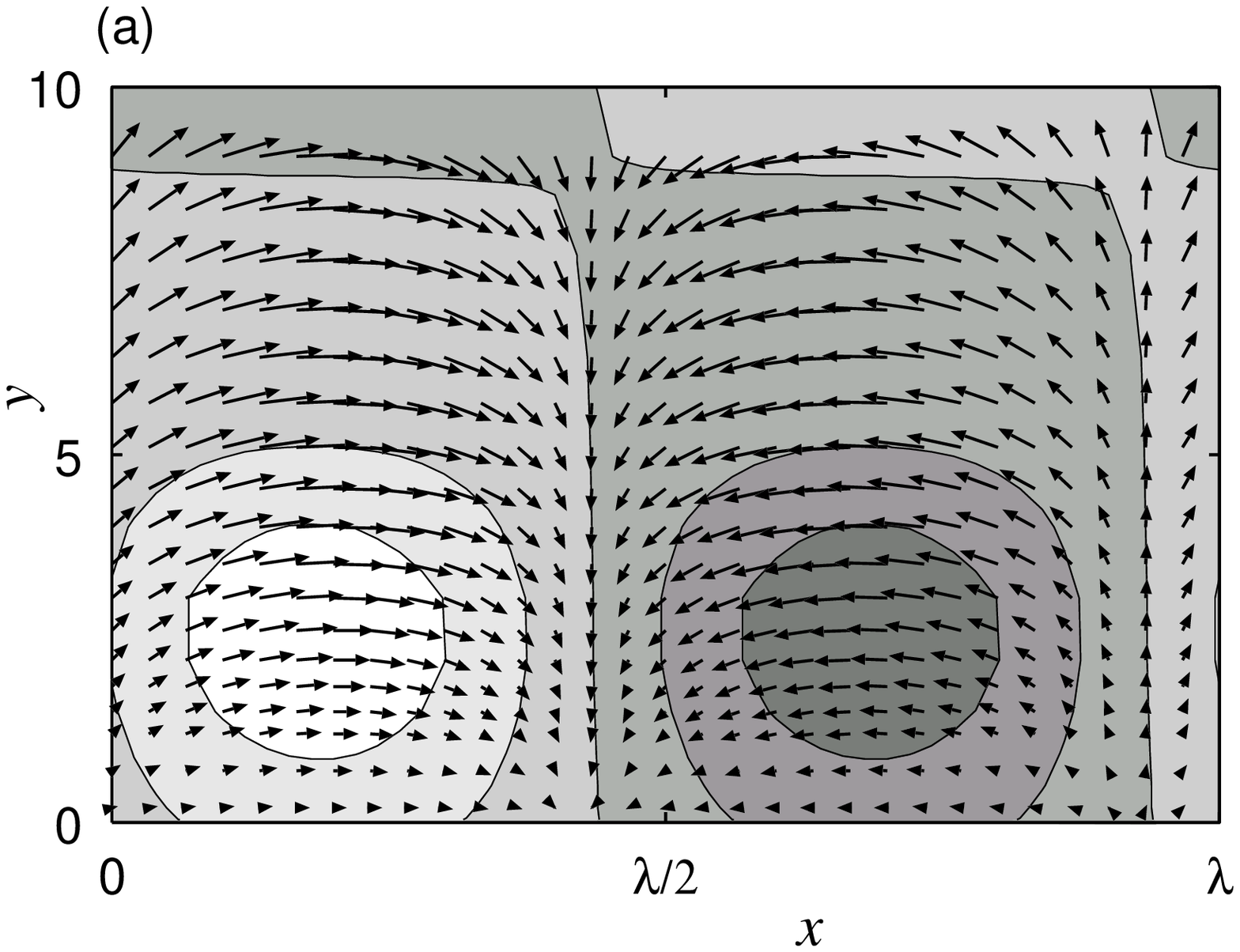}
\includegraphics[width=0.47\textwidth]{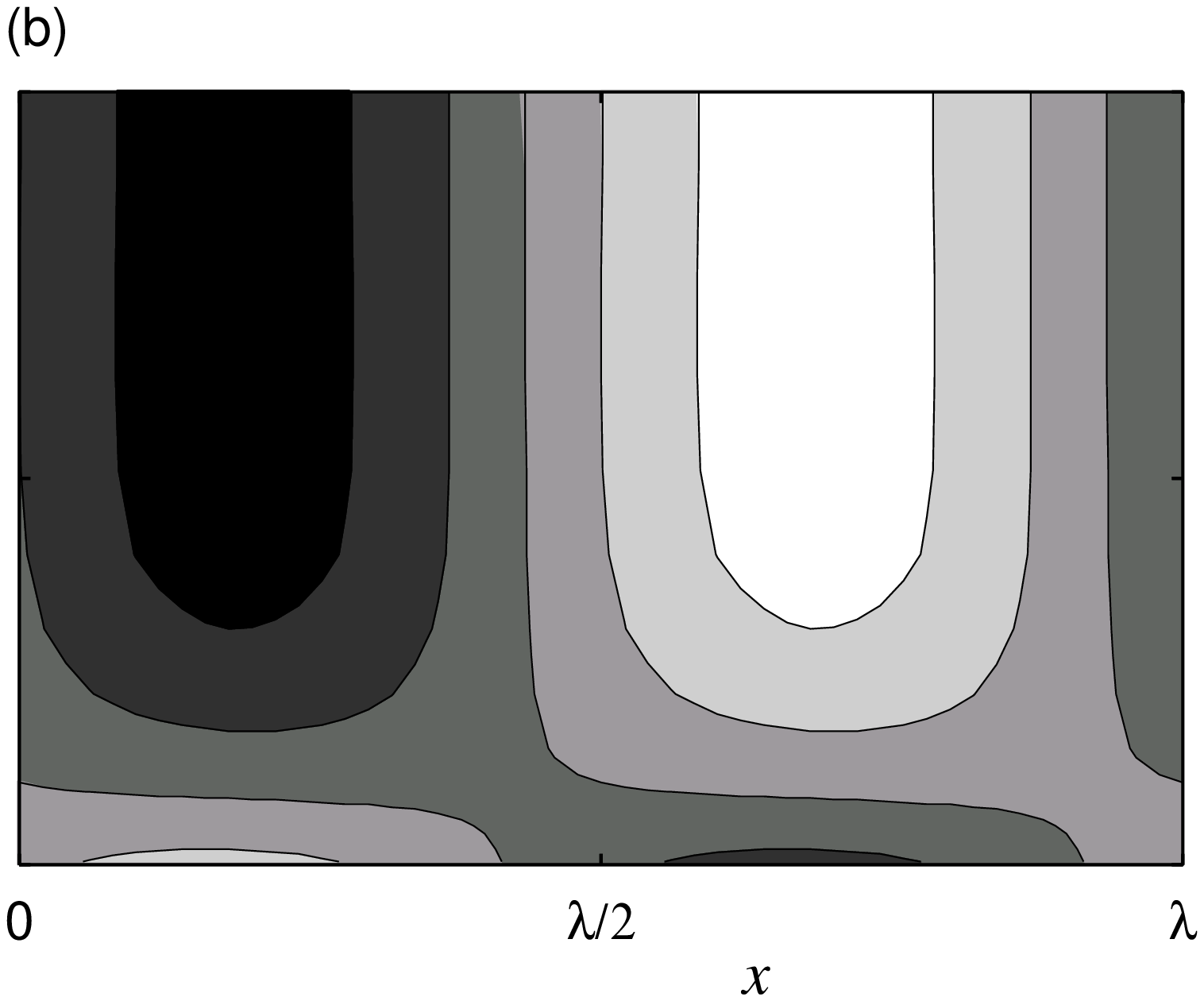}
\includegraphics[width=0.4\textwidth]{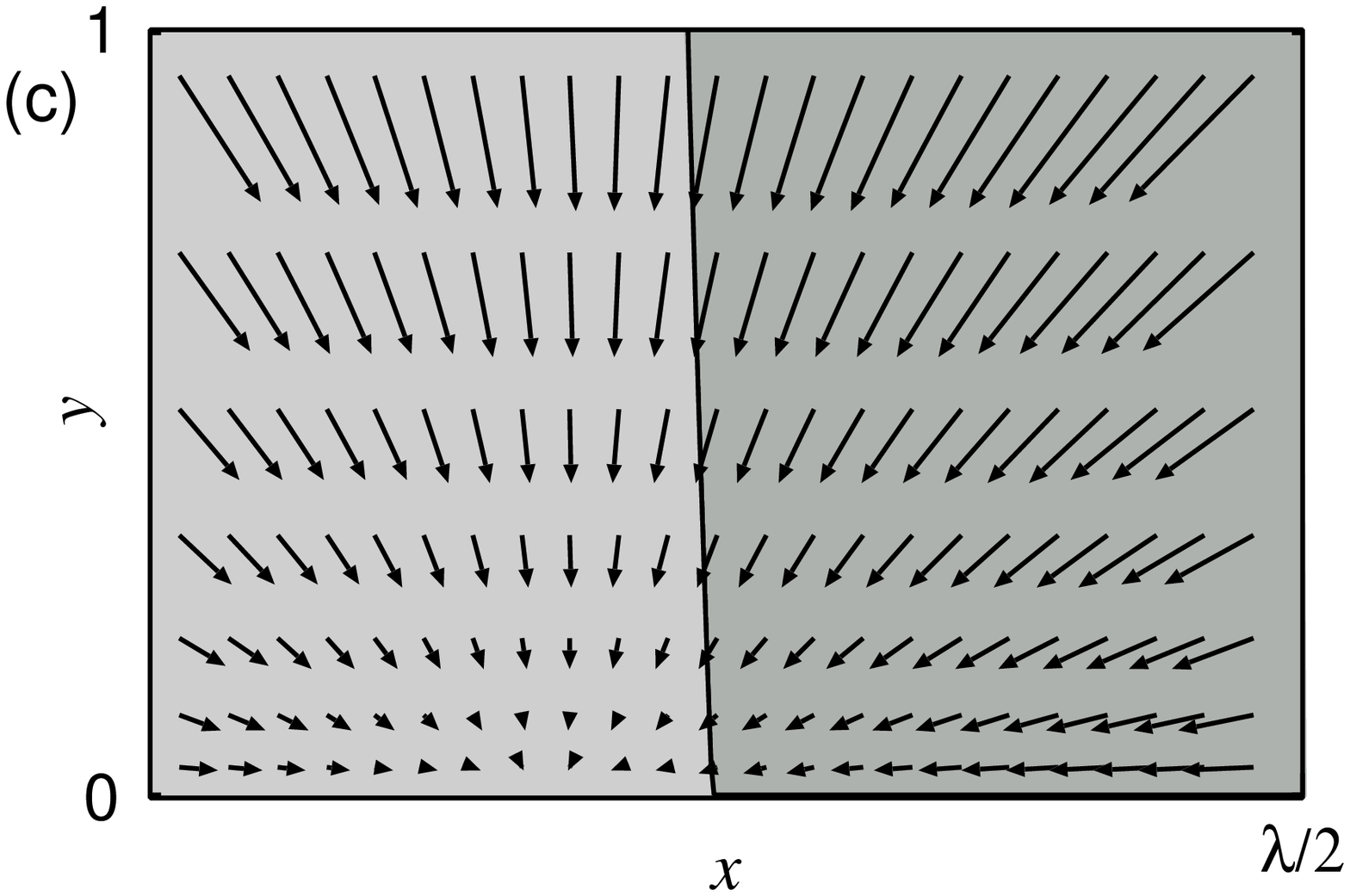}
\end{center}
\caption{The eigenfunctions of the unstable mode 
for $\phi=0.10$, $\theta=20^\circ$, and $\nu_0(0)=0.17$
(Q3 in figure \ref{phase010}),
whose wave number is $k=0.002$.
The contour of the density eigenfunction and 
the velocity eigenfunction in (a) 
show that the grains flow 
into the the region of the positive density perturbation.
In (b), the contour of the temperature perturbation is shown.
(c) is the magnification of (a) around $x\approx \lambda/2$, 
$y\approx 1$.
}\label{eigenu01}
\end{figure}

The unstable eigenmodes for $\theta=20^\circ$ and $\nu_0(0)=0.17$ 
(Q3 in figure \ref{phase010})
with $k=0.002$ are shown in figure \ref{eigenu01}.
The density and 
the velocity eigenfunctions
indicate that the grains flow into the 
the region of positive density perturbation 
(see the region around $x\approx\lambda/2$, $y\approx 0$; 
the magnification is shown (c)), 
thus the density perturbation will grow.
The difference from the case in figure \ref{eigenu} is that 
the velocity perturbation at the floor ($y=0$)
is in the positive $x$ direction in the region of the positive 
density perturbation; namely, the particles flow faster in the region
where they get dense.

\section{Discussions}\label{discussion}
We have calculated the steady flow solutions and 
examined their linear stability under the 
longitudinal perturbation
in the cases of $\phi=0.05$ and $\phi=0.10$.
The linear stability analysis revealed  that 
there are three unstable regimes A, B, and C 
in the $(h,\bar \nu)$ plane in both of the cases.
The regimes A and B are at the small $\bar \nu$ region,
while the regime C is at the large $\bar \nu$ region.
The difference between the regimes A and B 
is the density dependence of the flow velocity;
the denser flow is faster in the regime A,
while the flow with lower density is faster in the regime B.
The regime C
lies within the region where the density profile is non-monotonic
in the case of $\phi=0.05$, 
while the regime C includes a part of the region of
the monotonic density profiles for $\phi=0.10$,
although the region of the regime C roughly corresponds with
that of the non-monotonic density profiles.
In all the regimes, the dispersion relation of 
the unstable mode shows the features that
(i)$\alpha(0)=0$ and (ii)Re$[\alpha(k)]\propto k^2$
for small $k$.
The obtained unstable eigenmodes 
suggest that the instability causes the density wave.

In this section, we give some discussions on the mechanism
of the instability, comparison with simulations,
and the relationship with the transverse instability studied
by \cite{FP02}.

\subsection{Mechanism of the instability} \label{mechanism}
\subsubsection{Kinematic wave and the long-wave instability}
The long-wavelength instability which 
results in the density wave is well known for
quasi-one dimensional flows,
such as the wave formation in the film flow 
(\cite[Smith 1993]{S93};\cite[Ooshida 1999]{O99}),
the density wave formation in the granular flow in a narrow vertical pipe
(\cite[Raafat, Hulin \& Herrmann 1996]{RHH96};
\cite[Moriyama \etal 1998]{MKMH98}),
and the jam formation in the traffic flow 
(\cite[Kerner \& Konh\"auser 1993]{KK93};
\cite[Bando \etal 1995]{BHNSS95}; 
\cite[Mitarai \& Nakanishi 2000a, 2000b]{MN00,MN00-2}).
This instability is closely related to 
the continuity of ``density'' $\bar \rho$
(the thickness in the case of incompressible fluid,
the density per unit length along the pipe for the granular flow,
or the density of cars for the traffic flow):
$\bar \rho$ obeys the equation of continuity in the form
\begin{equation}
\frac{\partial \bar \rho(x,t)}{\partial t}+
\frac{\partial Q(x,t)}{\partial x}=0,
\label{kinematic0}
\end{equation}
where $Q$ is the flux.
The flux $Q$ may
be expressed in terms of $\bar \rho$ 
and its spatial derivatives, but
may also depend on time especially when the inertia effect exists.
In the long-wavelength and the long-time limit,  
the effect of the spatial derivatives of $\bar \rho$ 
and the inertia on $Q$ may be neglected. 
Then the flux is determined by the density;
$Q(x,t)=Q_0(\bar \rho(x,t))$, and 
equation (\ref{kinematic0}) becomes
\begin{equation}
\frac{\partial \bar \rho(x,t)}{\partial t}-
c\frac{\partial \bar \rho(x,t)}{\partial x}=0
\quad
\mbox{with}\quad
c=-\frac{\mbox{d} Q_0(\bar \rho)}{\mbox{d}\bar \rho}.
\end{equation}
The wave that can be described by this equation is called 
the kinematic wave (\cite[Whitham 1974]{W74}).
The effect of the inertia and spatial derivatives for the small but nonzero
wavelength appears as the growth rate quadratic in $k$ 
(\cite[Whitham 1974]{W74}; \cite[Ooshida 1999]{O99}), 
which causes the instability to yield the density wave.
The actual form of the flux $Q$
depends on a physical system, but we call the instability
caused through this mechanism 
``the long-wave instability'' after \cite{S93}.
Note that the unstable mode caused by the long-wave instability 
has the following three features 
for the complex growth rate $\alpha(k)$:
(i) $\alpha(0)=0$,
(ii) Re$[\alpha(k)]\propto k^2$ for small $k$,
and (iii) the phase velocity of the least stable mode
$c=\mbox{Im}[\alpha(k)]/k$
is given by $-\mbox{d}Q_0/\mbox{d}\bar \rho$ in the long-wavelength limit.

The unstable modes obtained in the present analysis of the two-dimensional
slope flow satisfy all of the features (i), (ii) and (iii);
the features (i) and (ii) have been already pointed out
in the text, and (iii) can be seen 
in figure \ref{dispmag}(b) 
in the case of $\phi=0.05$, $\nu_0(0)=0.05$, and $\theta=16^\circ$.
Actually, (iii) can be shown using (i) 
and the equation of continuity (\ref{eq:rho}) as shown in Appendix B.

These features strongly suggest that 
the longitudinal instability in the slope flow 
is the long-wave instability 
of the kinematic wave in a quasi-one dimensional system
for all of the regimes A, B, and C.

One may think that, in the regime C,
which lies in the region of the non-monotonic
density profile in the case of $\phi=0.05$ (figure \ref{phase005}),
the non-monotonic density profile might 
play a crucial role in the instability
as in the case of the transverse instability of 
Rayleigh-B\'ernard type,
where the convection occurs 
due to the non-monotonic density profile
(\cite[Forterre \& Pouliquen 2002]{FP02};
\cite[Carpen \& Brady 2002]{CB02});
however, the fact that the regime C contains 
a part of the region of the monotonic density profile 
in the case of $\phi=0.10$ (figure \ref{phase010})
suggests that the shape of the density profile does not determine
the instability. 
It should be also noted that the present longitudinal instability
occurs at the long wavelength while the transverse instability
appears at the finite wavelength.

\subsubsection{One dimensional model}
In order to confirm that
the longitudinal instability is the long-wave instability 
of the one-dimensional kinematic wave,
we now try to reduce our two-dimensional model
into the one-dimensional model that preserve
the major features of the original model.
In spite of crudeness of our procedure, we will see
the obtained one-dimensional model shows roughly the
same stability diagram for the long-wave 
instability of the kinematic wave.

The way we obtain the one-dimensional model is 
to integrate the original equations in $y$-direction
from $0$ to $\infty$;
the idea is similar to \cite{VP98},
where the one-dimensional model has been obtained for the flow in a
vertical chute by integrating the equations across the chute width. 

We define the one-dimensional density $\bar \rho(x,t)$,
the average velocity $\bar u(x,t)$, and the average temperature
$\bar T(x,t)$ as 
\begin{eqnarray}
\bar \rho(x,t)&\equiv&\int^{\infty}_{0}\nu(x,y,t)\mbox{d}y,\\
\bar \rho (x,t)\bar u(x,t)&\equiv&\int^{\infty}_{0}\nu(x,y,t) 
u(x,y,t)\mbox{d}y,\\
\bar \rho (x,t)\bar T(x,t)&\equiv&\int^{\infty}_{0}\nu(x,y,t) 
T(x,y,t)\mbox{d}y.
\end{eqnarray}

The one-dimensional equation of continuity 
obtained by integrating (\ref{eq:rho}) is
\begin{equation}
\partial_t\bar \rho+\partial_x(\bar \rho \bar u)=0,
\label{conti-one}
\end{equation}
where $\partial_t$ and $\partial_x$
represent 
$\partial/\partial t$ and $\partial/\partial x$,
respectively.  (\ref{conti-one}) is in the form
of (\ref{kinematic0}) with
the flux $Q(x,t)$ given by $Q=\bar \rho\bar u$.

By integrating the $x$-component of the equation of motion (\ref{eq:u}),
we obtain
\begin{eqnarray}
&&\partial_t (\bar \rho \bar u)+\partial_x 
\int^{\infty}_{0}\nu(x,y,t) u(x,y,t)^2 \mbox{d}y \nonumber \\
&&=\bar \rho \sin\theta+\Sigma_{yx}(\nu(0),T(0),u(0))
-\partial_x\left (\int^{\infty}_{0} \Sigma_{xx}(\nu, T,u)
\mbox{d}y\right),
\label{onedmotion}
\end{eqnarray}
which determines the evolution of $Q=\bar \rho \bar u$.
The second term in RHS is the shear stress at the floor and comes from 
the integration of $\partial_y \Sigma_{yx}$.

To simplify (\ref{onedmotion}), we make the
following approximations;
(i) Replace the second term in LHS
$\int^{\infty}_{0}\nu(x,y,t) u(x,y,t)^2 \mbox{d}y$
by $\bar \rho \bar u^2$,
(ii) Neglect the velocity in the $y$-direction, $v$,
and use the dilute limit expressions for 
the the pressure and viscosities
in $\Sigma_{xx}$,
namely, we replace
$\Sigma_{xx}$ by
$\nu T- (4/3)f_2(0)\sqrt{T}\partial_x u$, where
the first term comes from the pressure 
and the second term comes from the dilute limit of the 
shear viscosity.
The second viscosity $\zeta(\nu,T)$ is neglected 
because it is a higher order quantity in $\nu$ (see (\ref{eq:pmuetc})).
(iii) Estimate the integration of 
the shear viscosity term
by multiplying the integrand by the decay length of the density 
$(T/\cos\theta)$, and 
replacing $T$ by $\bar T$, $u$  by $\bar u$, 
i.e., 
replace $(4/3)f_2(0)\int^{\infty}_{0} (\sqrt{T}\partial_x u) \mbox{d}y$ by
$H(\bar T)\cdot (4/3)f_2(0) (\sqrt{\bar T}\partial_x \bar u) $,
where $H(\bar T)=\bar T/\cos \theta$. 

Then we obtain the equation
\begin{equation}
\partial_t (\bar \rho \bar u)+\partial_x (\bar \rho \bar u^2)
=\bar \rho \sin\theta+\Sigma_{yx}(\nu(0),T(0),u(0))
-\partial_x(\bar \rho \bar T)
+ (4/3)f_2(0)\partial_x (H(\bar T)\sqrt{\bar T}\partial_x \bar u).
\label{onedmotion2}
\end{equation}
For the shear stress at the floor $\Sigma_{yx}(\nu(0),T(0),u(0))$,
we have the boundary condition of momentum balance (\ref{eq:bc1}), i.e.
$\Sigma_{yx}(\nu(0),T(0),u(0))=-\eta^*(\nu(0),T(0))u(0)$.

To close equations (\ref{conti-one}) and (\ref{onedmotion2}),
we need the relation between ($\nu(0),T(0),u(0)$)
and ($\bar \rho,\bar T, \bar u$), and
the equation for the average temperature $\bar T$.
We simply assume that 
$u(0)=\bar u$ and $T(0)=\bar T$,
and we use the empirical relation between $\nu(0)$ and $\bar \rho$
for the steady flows; namely, 
we take $\nu(0)=F(\bar \rho)$, with the form of $F(\bar \rho)$ 
determined from the steady solution obtained numerically
for a fixed $\theta$ and different values of $\nu_0(0)$
by using $\nu_0(0)=F(\bar \rho_0)$ with 
$\bar \rho_0=\int_0^{\infty}\nu_0(y) \mbox{d}y$.
We further assume that $\bar T$ is also determined by 
the one-dimensional density $\bar \rho$, 
rather than to use the integrated equation for the temperature.
The form of  $\bar T=\bar T(\bar \rho)$ is 
determined from the steady flows, namely,
we assume 
$\bar T_0=(1/\bar \rho_0)\int^{\infty}_0\nu_0(y)T_0(y)
\mbox{d}y=\bar T(\bar \rho_0)$.

Now we finally obtain the one-dimensional model in the following
form;
\begin{eqnarray}
\partial_t\bar \rho+\partial_x(\bar \rho \bar u)&=&0,\\
\bar \rho \left[
\partial_t \bar u+\bar u \partial_x \bar u\right]
&=& a(\bar \rho)\left[U(\bar \rho)-\bar u\right]
-\partial_x(\bar \rho \bar T(\bar \rho )) \nonumber \\
&&+(4/3) f_2(0)\partial_x \left(H(\bar T(\bar \rho))
\sqrt{\bar T(\bar \rho)}\partial_x \bar u \right),
\label{onedv}
\end{eqnarray}
where 
\begin{equation}
a(\bar \rho)=\eta^*(F(\bar \rho), \bar T(\bar \rho)),\quad
U(\bar \rho)=\frac{\bar \rho\sin\theta}{a(\bar \rho)}.
\label{oneaandu}
\end{equation}

In this model, the steady solution is given by
$\bar\rho=\rho_0=\mbox{const.}$ 
and $\bar u=U(\rho_0)$,  
and the flux of the steady solution 
is given by $q_0=\rho_0 U(\rho_0)$.
The equation (\ref{onedv}) with (\ref{oneaandu})
shows that the velocity $U(\rho_0)$ is determined by
the balance between the acceleration by the gravity and 
the drag force from the floor.

This model is almost the same form as the traffic flow model 
proposed by \cite{KK93}, but with the different
form of the function $U(\bar \rho)$.
In the traffic flow model, $U(\bar \rho)$ 
is often called the ``optimal velocity function'',
which defines the density dependence of the car velocity
and is usually a decreasing function.
On the other hand, in the case of the granular flow on a slope, 
$U(\bar \rho)$ depends on the inclination angle 
and the boundary condition at the floor, and 
can take various forms as can be seen in
the plot of $\bar \rho_0$ vs $Q_0\approx q_0=\bar \rho_0 U(\bar \rho_0)$ 
\footnote{The flux $q_0$ in the one-dimensional model 
is not exactly the same as the flux $Q_0$
in the original two-dimensional flow,  
due to the approximations used to derive $U(\bar \rho)$. 
}
in figures \ref{steadyphase005}(b) and \ref{steadyphase010}(b).

The linear stability analysis of 
the steady solution can be performed analytically 
(for the traffic flow model, see e.g. \cite[Wada \& Hayakawa 1998]{WH98}).
It is easy to show that
the instability condition Re$[\alpha]>0$
yields
\begin{equation}
(a(\rho_0) U'(\rho_0))^2k^2>
\left(\frac{a(\rho_0)}{\rho_0}+\bar \mu(\rho_0) k^2 
\right)^2(\rho_0\bar T(\rho_0))'k^2,
\label{condition} 
\end{equation}
where $\bar\mu(\rho_0)\equiv (4/3) f_2(0)H(\bar T(\rho_0))
\sqrt{\bar T(\rho_0)}$ and 
the prime represents the differentiation by $\rho_0$.
The $k\to 0$ limit of (\ref{condition}) 
gives the stability criterion, 
\begin{equation}
(\rho_0 U'(\rho_0))^2>(\rho_0\bar T(\rho_0))'.
\label{criterion}
\end{equation}
The explicit form of the long-wavelength
expansion of the dispersion relation for the 
least stable mode is given by
\begin{equation}
\alpha(k)=-(\rho_0 U(\rho_0))'ik+
\frac{\rho_0}{a(\rho_0)}\left[
\left(\rho_0 U'(\rho_0)\right)^2-\left(\rho_0 
\bar T(\rho_0)\right)'\right]k^2+O(k^3),
\label{disp-oned}
\end{equation}
and we see that
the phase velocity of this mode in the long-wavelength limit
is given by $c=-(\rho_0 U(\rho_0))'
=-dq_0/d\rho_0$, which shows that this is the 
kinematic wave.
The instability arises when 
the coefficient of the $k^2$ term becomes positive,
which occurs if the change of the velocity against density fluctuation 
is too fast compared to the effect of the pressure
that reduces the density fluctuation.
Note that the criterion of the instability (\ref{criterion})
does not depend on the shear viscosity term $\bar \mu(\rho_0)$ 
because it only appears in the fourth order term in $k$
(see (\ref{condition}));
the approximation for that term does not crucially affect on the criterion.

\begin{figure}
\includegraphics[width=0.4\textwidth]{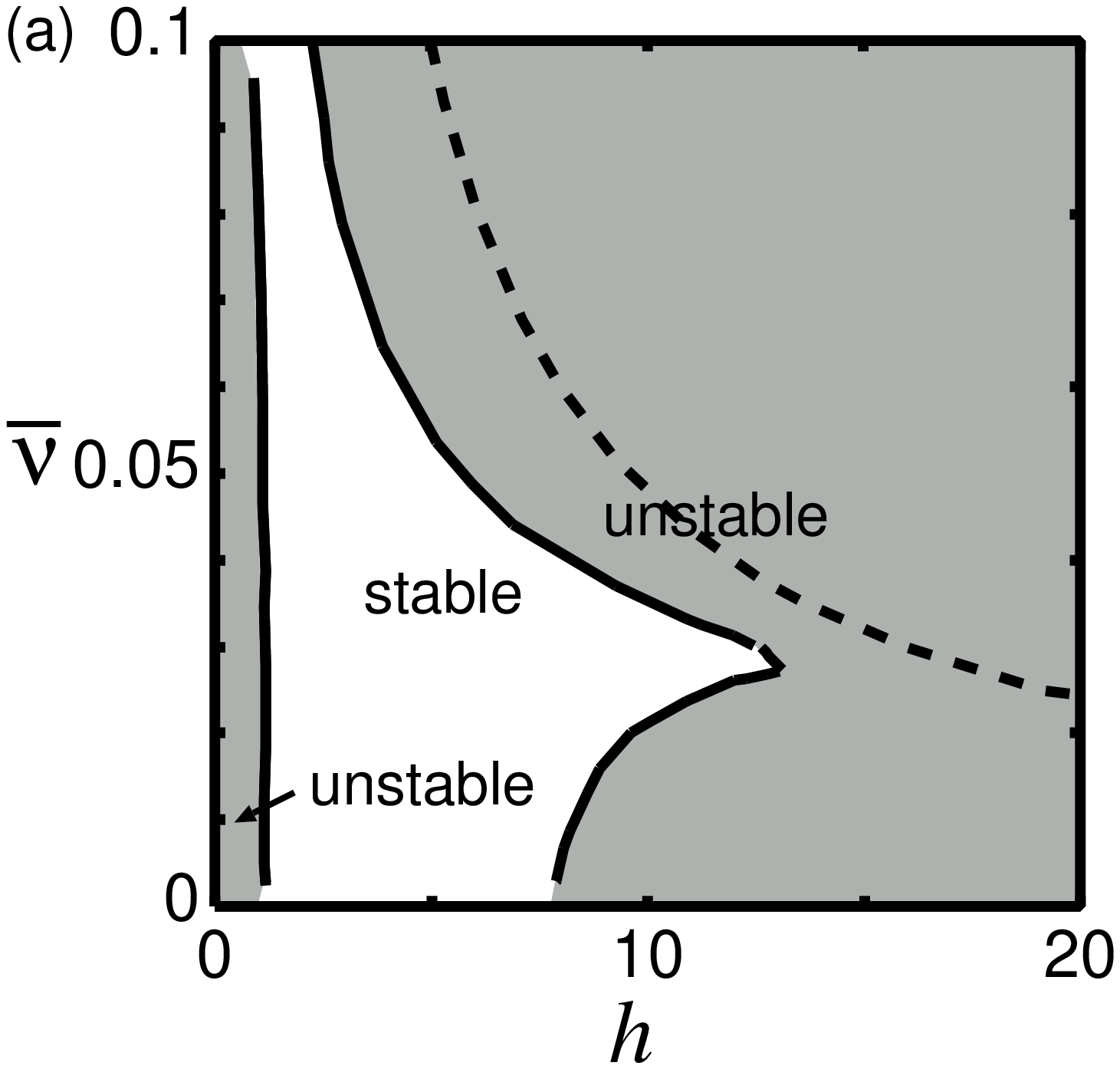}\hfill
\includegraphics[width=0.4\textwidth]{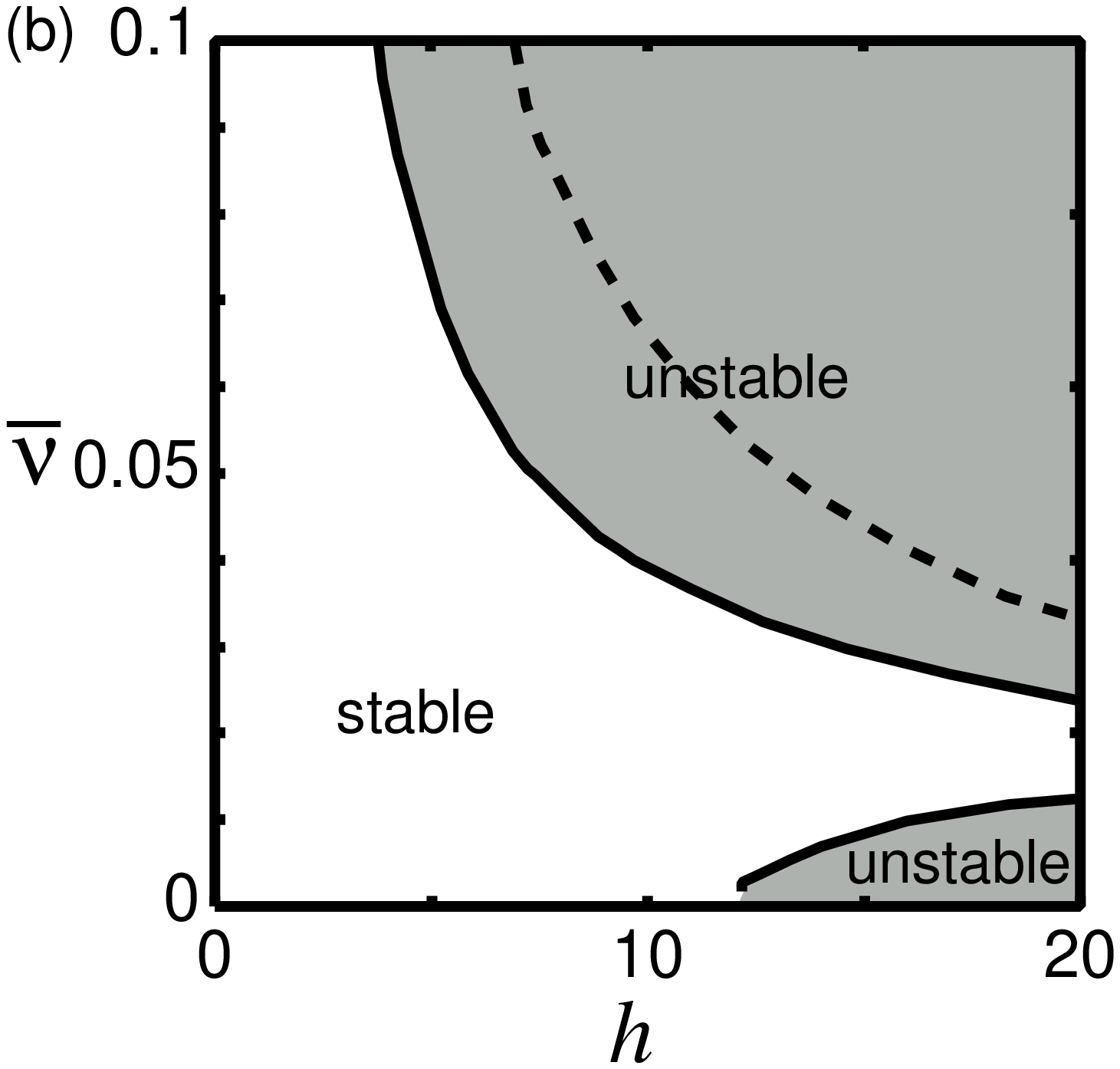}
\caption{The stability diagram for (a)$\phi=0.05$ and (b)$\phi=0.10$
obtained from the one-dimensional model.
The unstable (stable) regimes are shown by grey (white) regions.
The dashed line shows the boundary between 
the region of the non-monotonic density 
profiles and that of the monotonic density profiles for the original model.
}
\label{phaseoned}
\end{figure}

The stability diagrams for $\phi=0.05$ and $\phi=0.10$
obtained from this one-dimensional model
are shown in figures \ref{phaseoned}.
In spite of the crude approximations 
used in the derivation of the one-dimensional model,
the stability diagrams 
are qualitatively similar to those of the original model.
The similarity further indicates that
the density wave formation can be understood by
the long-wave instability 
in quasi-one dimensional systems, 
like in the film flow and the traffic flow.

It should be noticed that, in this one-dimensional model, 
the effect of the parameters in the original model 
such as $e_p$, $\phi$, and $\Phi$ are
included more or less implicitly in the functional forms of
$U(\bar \rho)$ and $\bar T(\bar \rho)$
($\phi$ also appears explicitly in (\ref{oneaandu}) 
through $\eta^*$), and
the $y$-dependences of the variables also 
affect $U(\bar \rho)$ and $\bar T(\bar \rho)$
through the integration.
Any changes that affect $U$ and $\bar T$ 
result in the changes of
the unstable regions (see the criterion (\ref{criterion})),
although the nature of the instability
remains the same.

Before concluding this subsection,
let us make a few comments on the works by
\cite[Wang, Jackson, \& Sundaresan(1997)]{W97} 
and \cite{VP98} on the stability analysis of granular flow in 
a vertical chute using hydrodynamic models of rapid granular flow; 
\cite{W97} performed the linear stability analysis numerically as 
in the present work,
and \cite{VP98} analyzed the density wave 
by deriving a one-dimensional model from 
hydrodynamic equations for rapid flow.

In the analysis of \cite{W97},
a parameter region has been found where the 
steady flow is unstable against longitudinal
long-wavelength perturbation to form the 
density wave (figure 9 and 10 in \cite[Wang {\em et~al.}\/1997]{W97}). 
This instability might be also understood as the long-wave instability 
observed in the present analysis.
On the other hand, they also found the instabilities
for finite wavelength perturbations,
which has not been observed here.

The analysis by \cite{VP98} shows clear similarity of the instability 
in the chute flow to the one in the slope flow.
The one-dimensional model they obtained 
(we call it the VP model)
has the mathematical structure and physical mechanism 
similar to those of our one-dimensional model:
the velocity of the steady solution is determined by the 
balance of the gravitational acceleration and the drag force from
the wall, and shows the long-wave instability
for which the criterion is determined by
the change of the velocity against density fluctuation 
and the pressure term.

These results suggest that the instabilities in
the vertical chute flow and the slope flow are in the same class.

\subsection{Comparison with the molecular dynamics simulations}
We find some qualitative agreements between the present results 
and our previous simulations 
(\cite[Mitarai \& Nakanishi 2001]{MN01}) as follows.

Our simulations were performed 
for a fixed inclination angle and a particular roughness of the slope
with the periodic boundary condition imposed along the flow
direction.
Within the examined parameter region,
the steady flow shows the monotonic density profile
(\cite[Mitarai, Hayakawa, \& Nakanishi 2002]{MHN02}),
and the flow with lower density has higher velocity.

It has been demonstrated that the density wave appears
only in the long system with low enough particle density.
We have performed the simulations for several sets of the 
slope length $L$ and the particle number $N$. 
In the case of the particle density 
$N\sigma/L\approx 1.0$ (single layer), 
the clear density wave is not formed
for $L=250.5\sigma$ and $L=501\sigma$, whereas the density wave appears
for $L=1002\sigma$. 
Upon changing the density with
a fixed system length $L=501\sigma$, the density wave is formed 
when $N\sigma/L\approx 0.75$, while the steady flow 
is stable for the denser cases with $N\sigma/L\approx 1.0$ and $2.0$. 

These tendencies of the simulations 
agree with the behaviors around the unstable regime B 
of the present model on the three points:
(i)The flow with lower density is faster,
(ii)The flow with lower density is less stable,
and (iii) The critical wavelength for instability is very long.

Regarding (iii), the critical wave length 
$\lambda_c=2\pi/k_c$  is much longer than the particle diameter;
$\lambda_c\approx 900\sigma$ 
for $\theta=16^\circ$ and $\nu_0(0)=0.05$, for example.
This seems to be comparable with 
our simulation results, where 
the critical slope length $L_c$ was between $500\sigma$ and $1000\sigma$
for $N\sigma/L\approx 1.0$. 
We do not understand yet how such a small wave number arises, 
but we suspect that it comes from 
the long mean free path in the large $y$ region where the density is low,
namely, the particles flying over the clusters 
for a long distance prevent the growth of clusters
in a smaller length scale.

Based on the observations (i), (ii), and (iii),
the parameters that we have simulated
happens to be in the regime B, but 
if the simulations are performed
with different densities,  inclination angles,
and/or boundary conditions at the floor, 
the behavior that corresponds to the regime A or C
may be found.
\subsection{Comparison with the stability analysis against
the transverse perturbations}\label{trans}
\cite{FP02} examined the linear stability of 
the granular flow on a wide slope against the perturbation
transverse to the flowing direction,
in order to understand the regular streak pattern 
along the flow direction observed in their experiments 
(\cite[Forterre \& Pouliqen 2001]{FP01}).
They mainly focused on the parameter region where the 
non-monotonic density profile 
(or what they call the ``inverted density profile'')
is observed, because
they expected that such a flow would be unstable
to form the vortex rolls 
from the analogy to the Rayleigh-B\'ernard instability.
They have shown that, 
the flow is unstable against the
transverse perturbation in the large part of the
parameter region where the inverted density profile is observed.
The unstable mode shows the vortex-like pattern,
and they concluded that the streaks observed in the experiments
result from the rolls of vortices.
They also found that the flow with the monotonic density profile
becomes unstable for some parameters,
but the detail has not been reported.

One of the differences between 
this transverse Rayleigh-B\'ernard type instability
and the longitudinal long-wave instability 
appears in the length scale of the instability.
The longitudinal instability occurs
against the long-wavelength perturbations
in the long-time behavior as can be 
seen in the dispersion relations,
while the transverse instability 
occurs at a finite length scale comparable with 
the vortex roll.
Our analysis shows that
there is a parameter region where both of the instability may
occur, around the region of 
the non-monotonic density profiles (the regime C).
It should be interesting to investigate how
the two instabilities interfere or not by the full three 
dimensional analysis.

In the large inclination angle,
\cite{FP02} also observed the square lattice pattern. 
This phenomenon cannot be understood by the simple superposition of 
the long-wave instability and the Rayleigh-B\'ernard type instability,
because  the length scale of the long-wave instability
is much longer than the one of the lattice pattern.

\section{Summary}\label{sum}
The steady flows and their
linear stability are analyzed for the granular flow on a slope
using the hydrodynamic model
with the constitutive relations
derived from the kinetic theory of inelastic spheres.
We mainly focused on the relatively low density region
where the density decays monotonically.

The stability diagram shows the
three unstable regimes A, B, and C in the both cases of
$\phi =0.05$ and $\phi=0.10$.
Two of the unstable regimes A and B
are in the lower density region,
and the regime C is in the high density region.
The difference between the regimes A and B 
is that, the denser flow is faster in the regime A
at the small $h$ side,
while the flow with lower density is faster in the regime B
at the large $h$ side.
The regime C is in the large $h$ and large $\bar \nu$ region;
it lies within the region of the non-monotonic density 
profile in the case of $\phi =0.05$, while 
it contains a part of the region of 
monotonic density profiles in the case of $\phi =0.10$,
although the region of the regime C roughly corresponds with
that of the non-monotonic density profile.
In all regimes, the instability occurs for the long-wavelength 
perturbations and results in the formation of density wave.
It has been found that the behaviors around the unstable regime B
agree with the tendencies in our previous simulations of 
density wave formations.

The dispersion of the complex growth rate $\alpha(k)$ has
the features that
(i)$\alpha(0)=0$, (ii)Re$[\alpha(k)]\propto k^2$ for small $k$,
and (iii)Im$[\alpha(k)]/k=-\mbox{d}Q_0/\mbox{d}\bar \rho_0$ in 
the long-wavelength limit.
These strongly suggest that the instability is the long-wave instability
of the kinematic waves, which is often 
found in quasi-one dimensional flows.
This is different from the transverse instability
studied by \cite{FP02}, where the flow is unstable at the finite 
wave number.

In order to confirm that the instability
is the long-wave instability of the kinematic wave,
we simplified the original equations rather heuristically
into a one-dimensional model, and 
showed that the long-wave instability occurs in the 
derived one-dimensional
model. The stability diagram obtained from the one-dimensional model
corresponds qualitatively to the one obtained
from the original equations.

\begin{acknowledgments}
N. M. is grateful to Ooshida Takeshi for informative discussions.
This work was partially supported by 
Hosokawa powder technology foundation
and Grant-in-Aid for JSPS fellows.
\end{acknowledgments}

\appendix
\section{Linearized equations}\label{lineareq}
In this appendix,
the linearized governing equations 
for the longitudinal perturbations
(\ref {eq:per1})-(\ref {eq:per3}) 
and the boundary conditions are given 
(\cite[Forterre \& Pouliquen 2002]{FP02}, 
see also \cite[Alam \& Nott 1998]{AN98}).
The superscript $0$ denotes that the quantities are for steady solution.
The subscript $\nu$ and $T$ denotes the partial derivatives
by the variables;
$p^0_\nu=\partial p(\nu,T)/\partial \nu|_{\nu=\nu_0,T=T_0}$.
The subscript $y$ indicates the total differential with respect to $y$,
namely, $p^0_y=\mbox{d}p^0(\nu(y),T(y))/\mbox{d}y=
[\partial p(\nu,T)/\partial \nu|_{\nu=\nu_0,T=T_0}]\nu_{0,y}+$
$[\partial p(\nu,T)/\partial T|_{\nu=\nu_0,T=T_0}]T_{0,y}$, 
$\nu_{0,y}=\mbox{d}\nu_0(y)/\mbox{d}y$,
and so on.
The expressions of $p(\nu,T)$, $\mu(\nu, T)$, $\zeta(\nu,T)$,
$\kappa(\nu,T)$, and $\Gamma(\nu,T)$ are given in (\ref{eq:pmuetc}),
and $\xi(\nu,T)=\zeta-2\mu/3$.
Then, by inserting (\ref{eq:per1})-(\ref{eq:per3})
into (\ref{eq:rho})-(\ref{eq:temp}) through
(\ref{eq:rho0})-(\ref{eq:T0}), we obtain
the following expressions;
\begin{eqnarray}
&&[\partial_t+u_0\partial_x]\nu_1
+[\nu_0\partial_x] u_1
+[\nu_{0,y}+\nu_0\partial_y]v_1=0,\label{eq:leqnu}\\
&&[\sin\theta-p^0_\nu\partial_x
+u_{0,yy}\mu^0_\nu
+u_{0,y}\mu_{\nu,y}^0+u_{0,y}\mu_{\nu}^0\partial_y]\nu_1\nonumber\\
&+&[-\nu_0\partial_t-\nu_0u_0\partial_x
+(\xi^0+2\mu^0)\partial_x^2
+\mu_y^0\partial_y+\mu^0\partial_y^2] u_1\nonumber\\
&+&[-\nu_0 u_{0,y}+
\mu_y^0\partial_x+(\xi^0+\mu_0)\partial_x\partial_y
]v_1\nonumber\\
&+&[-p_T^0\partial_x+u_{0,yy}\mu_T^0
+u_{0,y}\mu_{Ty}^0+u_{0,y}\mu_T^0\partial_y
]T_1=0,\\
&&[-\cos\theta-p^0_{\nu y}-p^0_\nu\partial_y+u_{0,y}\mu^0_\nu\partial_x
]\nu_1\nonumber\\
&+&[
\xi^0_y\partial_x+\xi^0\partial_x\partial_y+
\mu^0\partial_x\partial_y
]u_1\nonumber\\
&+&[-\nu_0\partial_t-\nu_0u_0\partial_x
+\xi_y^0\partial_y+
\xi^0\partial_y^2
+2\mu_y^0\partial_y+2\mu_0\partial_y^2
+\mu_0\partial_x^2
]v_1\nonumber\\
&+&[-p_{Ty}^0-p_T^0\partial_y 
+u_{0,y}\mu_T^0\partial_x
]T_1=0,\\
&&[\kappa^0_{\nu y}T_{0,y}+\kappa^0_{\nu}T_{0,yy}
+\kappa^0_{\nu}T_{0,y}\partial_y-\Gamma^0_\nu
+u_{0,y}^2\mu_\nu^0]\nu_1\nonumber \\
&+&[-p^0\partial_x+2\mu^0 u_{0,y}\partial_y
]u_1\nonumber \\
&+&\left[
-\frac{3}{2}\nu_{0}T_{0,y}-p^0\partial_y
+2\mu^0 u_{0,y}\partial_x\right]
v_1\nonumber \\
&+&\left[-\frac{3}{2}\nu_{0}\partial_t-\frac{3}{2}\nu_{0}u_0\partial_x
+\kappa^0\partial_x^2
+\kappa_y^0\partial_y+\kappa^0\partial_y^2\right.\nonumber\\
&&\left.
+\kappa^0_{Ty}T_{0,y}+\kappa^0_{T}T_{0,yy}+\kappa^0_{T}T_{0,y}\partial_y
-\Gamma_T^0+u_{0,y}^2\mu_T^0
\right]T_1=0.\label{eq:leqT}
\end{eqnarray}

The linearized boundary conditions at $y=0$ 
are obtained from (\ref{eq:bc1}) and (\ref{eq:bc2}) as
\begin{eqnarray}
\partial_y u_1&=&\phi [u_0 f_{7,\nu}^0\nu_1+f_7^0u_1], \label{eq:bc0u}\\
\partial_y T_1&=&
-f_{6,\nu}^0\left[\frac{1}{3}\phi u_0^2-\frac{1}{2}
\Phi T_0\right]\nu_1
-f_6^0\left[\frac{2}{3}\phi u_0u_1
-\frac{1}{2}
\Phi T_1
\right].\label{eq:bc0T}
\end{eqnarray}

\section{The long-wavelength limit of the phase velocity}
We have obtained the dispersion relation of the 
least stable mode, $\alpha=\alpha(k)$,
and numerically found that it satisfies
\begin{equation}
\alpha(0)=0.
\end{equation}
In this appendix, we show that the phase velocity 
of this mode, $c=\mbox{Im}(\alpha(k))/k$, satisfies
\begin{equation}
c=-\frac{\mbox{d}Q_0}{\mbox{d}\bar \rho_0}
\label{eq:vp0}
\end{equation}
in the long-wavelength limit $(k\to 0)$, which suggests
the mode is the kinematic wave.
Here, $\bar \rho_0$ and $Q_0$ are the one-dimensional density and the flux 
of the steady solution, respectively, 
defined in (\ref{rho0q0}).

Note that the derivative $\mbox{d}Q_0/\mbox{d}\bar \rho_0$ is taken
within a family of solutions for a fixed inclination angle as below.
For a given inclination angle, the density at the floor $\nu_0(0)=\beta$
is a continuous parameter to specify a steady solution.
To express $\beta$ dependence explicitly, we rewrite the steady
solutions as 
\begin{eqnarray}
\nu(x,y,z,t)&=&\nu_0(y;\beta),\label{eq:rho0-2}\\
\boldsymbol u(x,y,z,t)&=&(u_0(y;\beta),0,0),\label{eq:u0-2}\\
T(x,y,z,t)&=&T_0(y;\beta).\label{eq:T0-2}
\end{eqnarray}
Then $dQ_0/d\bar \rho_0$ is given by
\begin{equation}
\frac{\mbox{d} Q_0}{\mbox{d} \bar \rho_0}
=\frac{\mbox{d} Q_0(\beta)/\mbox{d} \beta}
{\mbox{d} \bar \rho_0(\beta)/\mbox{d} \beta}
=\frac{
\int_{0}^{\infty}
[\partial
(\nu_0(y;\beta) u_0(y;\beta))/\partial \beta]\mbox{d}y
}{
\int_{0}^{\infty}[\partial
\nu_0(y;\beta)/\partial \beta]\mbox{d}y
}.
\label{eq:dqdh}
\end{equation}

In order to show (\ref{eq:vp0}),
let us first express $c$ by eigenfunctions of the mode.
By linearization of the equation of continuity (\ref{eq:rho})
using (\ref{eq:lin}),
we obtain the following expression: 
\begin{equation}
\alpha \hat \nu(y)=-ik\left(\nu_0(y;\beta)\hat u(y)+
u_0(y;\beta)\hat \nu(y)\right)-\partial_y(\nu_0(y;\beta)\hat v(y)).
\label{linearconti}
\end{equation}
Integrating (\ref{linearconti}) from 
$y=0$ to $\infty$, 
we have
\begin{equation}
\alpha \int_0^{\infty} \hat \nu(y)\mbox{d}y
=-ik\int_0^{\infty} \left[\nu_0(y;\beta)\hat u(y)+
u_0(y;\beta)\hat \nu(y)\right] \mbox{d}y.
\label{eq:omega}
\end{equation}
Here, $\hat \nu(y), \hat u(y), \hat v(y)$, and $\hat 
T(y)$ depend on the wavenumber $k$,
and we expands these functions with respect to $k$, i.e.,
\begin{eqnarray}
\hat \nu(y)= \hat \nu_0(y)+ik \hat \nu_1(y)+.... ,
&\quad &
\hat u(y)= \hat u_0(y)+ik \hat u_1(y)+.... , \\
\hat v(y)= \hat v_0(y)+ik \hat v_1(y)+.... ,
&\quad &
\hat T(y)= \hat T_0(y)+ik \hat T_1(y)+....,  
\end{eqnarray}
where $({\hat \nu_i(y)},{\hat u_i(y)},{\hat v_i(y)},{\hat T_i(y)})$
do not contain $k$.
The long-wavelength expansion of the
dispersion relation $\alpha=\alpha(k)$
which satisfies $\alpha(0)=0$
is obtained from (\ref{eq:omega})
when $\int_0^\infty \hat \nu_0(y)\mbox{d}y\ne 0$, and we have
\begin{equation}
\alpha(k)= -ik\frac{\int_0^\infty \left[\nu_0(y;\beta )\hat u_0(y)+
u_0(y;\beta )\hat \nu_0(y)\right] \mbox{d}y}
{\int_0^\infty \hat \nu_0(y)\mbox{d}y}
+O(k^2).
\label{eq:omega2}
\end{equation}
Namely, $c$ is given by
\begin{equation}
c=-\frac{\int_0^\infty \left[\nu_0(y;\beta )\hat u_0(y)+
u_0(y;\beta )\hat \nu_0(y)\right] \mbox{d}y}
{\int_0^\infty \hat \nu_0(y)\mbox{d}y}
\label{eq:vp2}
\end{equation}
in the long-wavelength limit.

Now all we have to do is to express
$\vec{\hat X_0}\equiv (\hat \nu_0, \hat u_0, \hat v_0,\hat T_0)$
with respect to the steady solution
$\vec{X_0}(y;\beta)\equiv (\nu_0(y;\beta), u_0(y;\beta), 
0,T_0(y;\beta))$.
Let us write
(\ref{eq:rho})-(\ref{eq:temp}) and the boundary conditions
in the matrix form:
\begin{equation}
B\frac{\partial \vec{X}}{\partial t}=\vec N(\vec X).
\end{equation}
Here, $\vec X=(\nu, u, v,T)$, $B$ is 
a constant matrix, and $\vec N$
is a nonlinear operator.
A steady solution $\vec{X_0}(y;\beta)$ satisfies
\begin{equation}
\vec N(\vec X_0(y;\beta ))=0,
\label{eq:teijo}
\end{equation}
therefore, from (\ref{eq:lin}),
$\vec{\hat X} =(\hat\nu, \hat u, \hat v, \hat T)$
satisfies
\begin{equation}
\frac{\partial \vec N(\vec X))}{\partial \vec X}
|_{\vec X=\vec X_0(y;\beta )}
\vec{\hat X}=\alpha(k) B \vec{\hat X},
\label{eq:hatx}
\end{equation}
and expanding (\ref{eq:hatx}) by $k$ with $\alpha(0)=0$,
we have
\begin{equation}
\frac{\partial \vec N(\vec X))}{\partial \vec X}|_{\vec X=\vec X_0(y;\beta )}
\vec{\hat X_0}=0
\label{eq:x01}
\end{equation}
for the lowest order of $k$.
On the other hand, differentiating (\ref{eq:teijo})
by $\beta $, we obtain
\begin{equation}
\frac{\partial \vec N(\vec X))}
{\partial \vec X}|_{\vec X=\vec X_0(y;\beta )}
\frac{\partial\vec{X_0}(y;\beta )}{\partial \beta }=0.
\label{eq:x02}
\end{equation}
It is plausible that the mode of zero eigenvalue for $k=0$ 
does not degenerate, 
because the mass is the only one conserved quantity
(the momentum is lost at the floor, and the energy is dissipated.).
Thus, from (\ref{eq:x01}) and (\ref{eq:x02}),
we obtain $\vec{\hat X_0}\propto
\partial \vec{X_0}(y;\beta )/\partial \beta$,
or more explicitly, 
\begin{equation}
(\hat \nu_0, \hat u_0, \hat v_0,\hat T_0)
\propto 
\left(\frac{\partial \nu_0(y;\beta)}{\partial \beta}, 
\frac{\partial u_0(y;\beta)}{\partial \beta}, 
0,\frac{\partial T_0(y;\beta)}{\partial \beta}\right).
\label{eq:last}
\end{equation}
Using (\ref{eq:vp2}) and (\ref{eq:last}), we get
\begin{equation}
c= -\frac{\int_0^\infty \left[\nu_0(y;\beta )(\partial{
u_0(y;\beta )}/\partial \beta )
+u_0(y;\beta )
(\partial{
\nu_0(y;\beta )}/\partial \beta )\right] \mbox{d}y}
{\int_0^\infty \left(\partial{\nu_0(y;\beta )}/\partial \beta \right)
\mbox{d}y},
\end{equation}
and by comparing this with (\ref{eq:dqdh}), we have
\begin{equation}
c=-\frac{\mbox{d}Q_0/\mbox{d}\beta}
{\mbox{d}\bar \rho_0/\mbox{d}\beta}
=-\frac{\mbox{d}Q_0}{\mbox{d}\bar \rho_0},
\end{equation}
which is (\ref{eq:vp0}).


\begin{thebibliography}{33}
\expandafter\ifx\csname natexlab\endcsname\relax\def\natexlab#1{#1}\fi

\bibitem[Ahn {\em et~al.\/}(1992)Ahn, Brennen \& Sabersky]{ABS92}
{\sc Ahn, H., Brennen, C.~E. \& Sabersky, R.~H.} 1992 Analysis of the fully
  developed chute flow of granular materials. {\em J. Appl. Mech.\/} {\bf 59},
  109--119.

\bibitem[Alam \& Nott(1998)]{AN98}
{\sc Alam, M. \& Nott, P.~R.} 1998 Stability of plane Couette flow of a
  granular material. {\em J. Fluid Mech.\/} {\bf 377}, 99--136.

\bibitem[Anderson {\em et~al.\/}(1999)Anderson, Bai, Bischof, Blackford,
  Demmel, Dongarra, Du~Croz, Greenbaum, Hammarling, McKenney \& Sorensen]{laug}
{\sc Anderson, E., Bai, Z., Bischof, C., Blackford, S., Demmel, J., Dongarra,
  J., Du~Croz, J., Greenbaum, A., Hammarling, S., McKenney, A. \& Sorensen, D.}
  1999 {\em {LAPACK} Users' Guide\/}, 3rd edn. Philadelphia, PA: Society for
  Industrial and Applied Mathematics.

\bibitem[Anderson \& Jackson(1992)]{AJ92}
{\sc Anderson, K.~G. \& Jackson, R.} 1992 A comparison of the solutions
of some proposed equations of motion of granular materials for
fully developed flow down inclined planes. {\em J. Fluid Mech.\/} 
{\bf 241}, 145--168.


\bibitem[Bando {\em et~al.\/}(1992)]{BHNSS95}
{\sc Bando, M., Hasebe, K., Nakayama, A., Shibata, A., 
\& Sugiyama, A.} 1995 Dynamical model of traffic congestion and 
numerical simulation. {\em Phys. Rev. E\/} {\bf 51},
  1035--1042.

\bibitem[Boyd(2001)]{boyd}
{\sc Boyd, J.~P.} 2001 {\em Chebyshev and Foulier Spectral Methods (Second
  Edition)\/}. Dover.

\bibitem[Campbell(1990)]{C90}
{\sc Campbell, C.~S.} 1990 Rapid granular flows. {\em Ann. Rev. Fluid Mech.\/}
  {\bf 22}, 57--92.

\bibitem[Canuto {\em et~al.\/}(1988)Canuto, Hussaini, Quarteroni \&
  Zang]{canuto}
{\sc Canuto, C., Hussaini, M.~Y., Quarteroni, A. \& Zang, T.~A.} 1988 {\em
  Spectral Methods in Fluid Dynamics\/}. Springer Series in computational
  physics.

\bibitem[Carpen \& Brady(2002)]{CB02}
{\sc Carpen, I.~C. \& Brady, J.~F.} 2002 
  Gravitational instability in suspension flow. 
{\em J. Fluid Mech.\/} {\bf 472}, 201--210.

\bibitem[Cundall \& Strack(1979)]{CS79}
{\sc Cundall, P.~A. \& Strack, O. D.~L.} 1979 A discrete numerical model for
  granular assemblies. {\em Geotechnique\/} {\bf 29}, 47--65.

\bibitem[Forterre \& Pouliquen(2001)]{FP01}
{\sc Forterre, Y. \& Pouliquen, O.} 2001 Longitudinal vortices in granular
  flows. {\em Phys. Rev. Lett.\/} {\bf 86}, 5886--5889.

\bibitem[Forterre \& Pouliquen(2002)]{FP02}
{\sc Forterre, Y. \& Pouliquen, O.} 2002 Stability analysis of rapid granular
  chute flows: formation of longitudinal vortices. {\em J. Fluid Mech.\/} {\bf
  467}, 361--387.

\bibitem[Goldhirsh(2003)]{G03}
{\sc Goldhirsh, I} 2003 Rapid granular flows. 
{\em Annu. Rev. Fluid Mech.\/} {\bf 35} 267--293.

\bibitem[Gottilieb {\em et~al.\/}(1984)Gottilieb, Hussaini \& Orsag]{spectral}
{\sc Gottilieb, D., Hussaini, M.~Y. \& Orsag, S.~A.} 1984 Theory and
  applications of spectral methods. In {\em Spectral Methods for Partial
  Differential Equations\/} (ed. R.~G. Voigt, D.~Gottilieb \& M.~Y. Hussaini).
  SIAM.

\bibitem[Jaeger {\em et~al.\/}(1996)Jaeger, Nagel \& Behringer]{JNB96}
{\sc Jaeger, H.~M., Nagel, S.~R. \& Behringer, R.~P.} 1996 Granular solids,
  liquids, and gases. {\em Rev. Mod. Phys.\/} {\bf 68}, 1259--1273.

\bibitem[Jenkins(1992)]{J92}
{\sc Jenkins, J.~T.} 1992 Boundary conditions for rapid granular flow: flat,
  frictional walls. {\em J. Appl. Mech.\/} {\bf 59}, 120--127.

\bibitem[Jenkins \& Richman(1986)]{JR86}
{\sc Jenkins, J.~T. \& Richman, M.~W.} 1986 Boundary conditions for plane flows
  of smooth, nearly elastic, circular disks. {\em J. Fluid Mech.\/} {\bf 171},
  53--69.

\bibitem[Jenkins \& Savage(1983)]{JS83}
{\sc Jenkins, J.~T. \& Savage, S.~B.} 1983 A theory for the rapid flow of
  identical, smooth, nearly elastic, spherical particles. {\em J. Fluid
  Mech.\/} {\bf 130}, 187--202.

\bibitem[Johnson \& Jackson(1987)]{JJ87}
{\sc Johnson, P.~C. \& Jackson, R.} 1987 Frictional-collisional constitutive
  relations for granular materials, with application to plane shearing. {\em J.
  Fluid Mech.\/} {\bf 176}, 67--93.

\bibitem[Johnson, Nott \& Jackson(1990)]{JNJ90}
{\sc Johnson, P.~C., Nott, P. \& Jackson, R.} 1990 
Frictional-collisional equations 
of motion for particulate flows and their application to chutes . 
{\em J. Fluid Mech.\/} {\bf 210}, 501--535.

\bibitem[Kerner \& Konh\"auser(1993)]{KK93}
{\sc Kerner, B.~S. \& Konh\"auser, P.} 1993 Cluster effect in initially
  homogeneous traffic flow. {\em Phys. Rev. E\/} {\bf 48}, R2335--R2338.

\bibitem[Louge \& Keast(2001)]{LK01}
{\sc Louge, M.~Y. \& Keast, S.~C.} 2001 On dense granular flows down flat
  frictional inclines. {\em Phys. Fluids\/} {\bf 13}, 1213--1233.

\bibitem[Lun \& Savage(1986)]{LS86}
{\sc Lun, C. K.~K. \& Savage, S.~B.} 1986 The effects of an impact velocity
  dependent coefficient of restitution on stresses developed by sheared
  granular materials. {\em Acta Mech.\/} {\bf 63}, 15--44.

\bibitem[Lun {\em et~al.\/}(1984)Lun, Savage, Jeffrey \& Chepurniy]{LSJC84}
{\sc Lun, C. K.~K., Savage, S.~B., Jeffrey, D.~J. \& Chepurniy, N.} 1984
  Kinetic theories for granular flow: inelastic particles in Couette flow and
  slightly inelastic particles in a general flowfield. {\em J. Fluid Mech.\/}
  {\bf 223}, 223--256.

\bibitem[Malik(1990)]{M90}
{\sc Malik, M.~R.} 1990 Numerical methods for hypersonic boundary layer
  stability. {\em J. Comput. Phys.\/} {\bf 86}, 376--413.

\bibitem[Mayer \& Powell(1992)]{MP92}
{\sc Mayer, E.~W. \& Powell, K.~G.} 1992 Viscous and inviscid instabilities of
  a trailing vortex. {\em J. Fluid Mech.\/} {\bf 245}, 91--114.

\bibitem[Mitarai {\em et~al.\/}(2002)Mitarai, Hayakawa \& Nakanishi]{MHN02}
{\sc Mitarai, N., Hayakawa, H. \& Nakanishi, H.} 2002 Collisional granular flow
  as a micropolar fluid. {\em Phys. Rev. Lett.\/} {\bf 88}, 174301.

\bibitem[Mitarai \& Nakanishi(2000{\natexlab{{\em a\/}}})]{MN00-2}
{\sc Mitarai, N. \& Nakanishi, H.} 2000{\natexlab{{\em a\/}}} Convective
  instability and structure formation in traffic flow. {\em J. Phys. Soc.
  Jpn\/} {\bf 69}, 3752--3761.

\bibitem[Mitarai \& Nakanishi(2000{\natexlab{{\em b\/}}})]{MN00}
{\sc Mitarai, N. \& Nakanishi, H.} 2000{\natexlab{{\em b\/}}} Spatiotemporal
  structure of traffic flow in a system with an open boundary. {\em Phys. Rev.
  Lett.\/} {\bf 85}, 1766--1769.

\bibitem[Mitarai \& Nakanishi(2001)]{MN01}
{\sc Mitarai, N. \& Nakanishi, H.} 2001 Instability of dilute granular flows on
  rough slope. {\em J. Phys. Soc. Jpn.\/} {\bf 70}, 2809--2812.

\bibitem[Mitarai \& Nakanishi(2003)]{MN03}
{\sc Mitarai, N. \& Nakanishi, H.} 2003 Hard-sphere limit of soft-sphere model
  for granular materials: Stiffness dependence of steady granular flow. {\em
  Phys. Rev. E\/} {\bf 67}, 021301.

\bibitem[Moriyama {\em et~al.\/}(1998)]{MKMH98}
{\sc Moriyama, O., Kuroiwa, N., Matsushita, M.
 \& Hayakawa, H.} 1998 4/3 law of granular particles flowing through
a vertical pipe. {\em Phys. Rev. Lett.\/} {\bf 80}, 2833--2836.

\bibitem[Ooshia (1999)]{O99}
{\sc Ooshida, T.} 1999 
Surface equation of falling film flows with moderate Reynolds number
and large but finite Weber number. {\em Phys. Fluids\/} {\bf 11}, 3247--3269.

\bibitem[Pouliquen(1999)]{P99}
{\sc Pouliquen, O.} 1999 Scaling laws in granular flows down rough 
inclined planes. {\em Phys. Fluids\/} {\bf 11}, 542--545.

\bibitem[Prasad {\em et~al.\/}(2000)Prasad, Pal \& R\"omkens]{PPR00}
{\sc Prasad, S.~N., Pal, D. \& R\"omkens, M. J.~M.} 2000 Wave formation on a
  shallow layer of flowing grains. {\em J. Fluid Mech.\/} {\bf 413},
  89--110.

\bibitem[Raafat {\em et~al.\/}(1996)Raafat, Hulin \& Herrmann]{RHH96}
{\sc Raafat, T., Hulin, J.~P. \& Herrmann, H.~J.} 1996 
 Density waves in dry granular media falling through a vertical pipe. 
{\em Phys. Rev. E\/} {\bf 53}, 4345--4350.

\bibitem[Richman(1988)]{R88}
{\sc Richman, M.~W.} 1988 Boundary conditions based upon a modified Maxwellian
  velocity distribution for flows of identical, smooth, nearly elastic spheres.
  {\em Acta Mechanica\/} {\bf 75}, 227--240.

\bibitem[Savage(1984)]{S84}
{\sc Savage, S.~B.} 1984 The mechanics of rapid granular flows.
  {\em Adv. Appl. Mech.\/} {\bf 24}, 289--366.

\bibitem[Smith(1993)]{S93}
{\sc Smith, M.K.} 199 
The mechanism for the long-wave instability in thin liquid films. 
{\em J. Fluid Mech.\/} {\bf 217}, 469--485.

\bibitem[Valance \& Pennec(1998)]{VP98}
{\sc Valance, A. \& Pennec, T.~L.} 1998 Nonlinear dynamics of density waves in
  granular flows through narrow vertical channels. {\em Euro. Phys. J. B\/}
  {\bf 5}, 223--229.

\bibitem[Wang {\em et~al.\/}(1997)]{W97}
{\sc Wang, C., Jackson, R. \& Sundaresan S.} 1997 
Instabilities of fully developed rapid flow of a granular material 
in a channel.
{\em J. Fluid Mech.\/} {\bf 342}, 179--197.

\bibitem[Wada \& Hayakawa(1998)]{WH98}
{\sc Wada, S. \& Hayakawa, H.} 1998 
Kink solution in a fluid model of traffic flow. {\em J. Phys. Soc. Jpn.\/}
  {\bf 67}, 763--766.


\bibitem[Whitham(1974)]{W74}
{\sc Whitham, G.B.} 1974 {\em Linear and nonlinear waves\/}. 
John Wiley \& Sons.
\end{thebibliography}
\end{document}